\documentclass[journal=jctcce,manuscript=article]{achemso}

\usepackage{chemformula} 
\usepackage[T1]{fontenc} 
\usepackage{xcolor}
\usepackage{bbm}
\usepackage{amsmath,amssymb}
\usepackage{bbold}
\usepackage{tabularx}
\definecolor{riverlane_green}{RGB}{0, 150, 143}
\definecolor{avired}{RGB}{216, 27, 96}
\usepackage{hyperref}
\hypersetup{
  colorlinks   = true, 
  urlcolor     = riverlane_green, 
  linkcolor    = riverlane_green, 
  citecolor   = avired  
}
\usepackage{braket}

\newcommand{\Tau}{\hat{\mathcal{T}}}

\newcommand{\tr}{ {\bf r}}
\newcommand{\tx}{ {\bf x}}

\newcommand{\tP}{ {\bf P}}

\newcommand{\tG}{ {\bf G}}

\newcommand{\INTr}[1]{\int\limits_{#1} d^3 \tr\,}
\newcommand{\IINTr}[1]{\iint\limits_{#1} d^3\tr \, d^3\tr'\,}

\newcommand{\adps}{\hat a^{\dagger}_{p\sigma}}

\newcommand{\aqs}{\hat a_{q\sigma}}

\AtBeginDocument{} 

\newcommand{\ka}{\kappa}
\newcommand{\core}{ {\text{core}}}

\author{Aleksei V. Ivanov}
\email{aleksei.ivanov@riverlane.com}
\affiliation{Riverlane Ltd, St Andrews House, 59 St Andrews Street, Cambridge, CB2 3BZ, United Kingdom}

\author{Andrew Patterson}
\affiliation{Riverlane Ltd, St Andrews House, 59 St Andrews Street, Cambridge, CB2 3BZ, United Kingdom}

\author{Marius Bothe}
\affiliation{Riverlane Ltd, St Andrews House, 59 St Andrews Street, Cambridge, CB2 3BZ, United Kingdom}
\alsoaffiliation{Astex Pharmaceuticals, 436 Cambridge Science Park, Cambridge, CB4 0QA, United Kingdom}

\author{Christoph S\"underhauf}
\affiliation{Riverlane Ltd, St Andrews House, 59 St Andrews Street, Cambridge, CB2 3BZ, United Kingdom} 

\author{Bjorn K. Berntson}
\affiliation{Riverlane Ltd, St Andrews House, 59 St Andrews Street, Cambridge, CB2 3BZ, United Kingdom}

\author{Jens Jørgen Mortensen}
\affiliation{CAMd, Department of Physics, Technical University of Denmark, 2800 Kgs. Lyngby, Denmark}

\author{Mikael Kuisma}
\affiliation{CAMd, Department of Physics, Technical University of Denmark, 2800 Kgs. Lyngby, Denmark}

\author{Earl Campbell}
\affiliation{Riverlane Ltd, St Andrews House, 59 St Andrews Street, Cambridge, CB2 3BZ, United Kingdom}
\alsoaffiliation{School of Mathematical and Physical Sciences, Sheffield S3 7RH, UK}

\author{R\'obert Izs\'ak}
\affiliation{Riverlane Ltd, St Andrews House, 59 St Andrews Street, Cambridge, CB2 3BZ, United Kingdom}

\title[UPAW]{Quantum Computation of Electronic Structure with Projector Augmented-Wave Method and Plane Wave Basis Set}

\begin{document}

\begin{tocentry}

\includegraphics[width=0.6\linewidth]{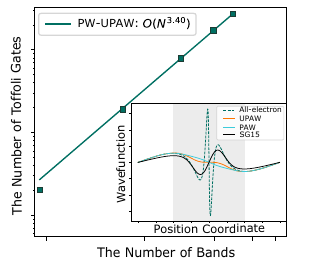}

\end{tocentry}

\begin{abstract}
Quantum simulation of materials is a promising application area of quantum computers. To practically realize this promise, we must reduce quantum resources while maintaining accuracy. In electronic structure calculations on classical computers, resource reduction has been achieved by using the projector augmented-wave method (PAW) and plane wave basis sets. However, the PAW method generalized for many-body states introduces non-orthogonality effects which impede its direct application to quantum computing.  In this work, we develop a unitary variant of the PAW (UPAW) that preserves the orthogonality constraints. We provide a linear-combination-of-unitaries decomposition of the UPAW Hamiltonian to enable ground state estimation using qubitized quantum phase estimation. Additionally, we further improve algorithmic efficiency by extending classical down-sampling techniques into the quantum setting.  We then estimate quantum resources for crystalline solids to estimate the energy within chemical accuracy with respect to the full basis set limit, and also consider a supercell approach which is more suitable for calculations of defect states. We provide the quantum resources for energy estimation of a nitrogen-vacancy defect centre in diamond which is a challenging system for classical algorithms and a quintessential problem in the studies of quantum point defects.
\end{abstract}

\maketitle

\section{Introduction}

Quantum computing for materials applications has made significant progress in recent years, driven not only by the development of novel quantum algorithms~\cite{abrams_simulation_1997,abrams_quantum_1999,ortiz_quantum_2001,aspuru-guzikSimulatedQuantumComputation2005a} but also by the adaptation of methodologies from classical quantum chemistry aimed to reduce the computational resources. These methodologies include, but are not limited to factorization techniques~\cite{berry_qubitization_2019,von_burg_quantum_2021,lee_even_2021}, the choice of different single-particle basis sets~\cite{babbush_low-depth_2018,song2023periodic,ivanov_quantum_2023,rubin_fault-tolerant_2023,Clinton2024} , and the use of Goedecker-Teter-Hutter norm-conserving pseudopotentials~\cite{goedecker_separable_1996,hartwigsen_relativistic_1998} in the first-quantization plane-wave formalism~\cite{su_yuan_fault-tolerant_2021,zini_quantum_2023, berry_quantum_2023}.
In classical simulations, it is well known that using ultrasoft pseudopotentials~\cite{Vanderbilt_soft_1990} or the projector augmented-wave~\cite{blochl_projector_1994} (PAW) method improves computational efficiency and accuracy~\cite{lejaeghere_reproducibility_2016,prandini2018precision,bosoni_how_2023}. Given this success, one might anticipate similar benefits in quantum computing. In this work, we will address the use of PAW for fault-tolerant quantum computation of energy states of an \textit{ab-initio} many-body Hamiltonian.

When applied to many-body calculations, the PAW approach has so far been employed as an all-electron method~\cite{taheridehkordi_phaseless_2023,humer_approaching_2022,gruneis_coupled_2015,gruneis_natural_2011,gruneis_second-order_2010,marsman_second-order_2009}, \textit{i.e.} the Hamiltonian acts on the full all-electron wavefunction space, typically with frozen core electrons. Representing such wavefunctions on a plane-wave or real-space grid, as done in some quantum algorithms, requires a large number of basis functions. In contrast, when applied to Kohn-Sham (KS) equations, the single-particle KS Hamiltonian is conjugated by the PAW transformation~\cite{blochl_projector_1994}, allowing the generalized KS equations to be solved for pseudo-orbitals represented by a much smaller set of plane waves. In our adaptation of PAW, we follow this latter strategy: First, we generalize the PAW transformation to many-electron wavefunctions, then perform a similarity transformation. This results in a generalized Schrödinger equation with non-orthogonal wavefunctions (a generalized eigenvalue problem). However, non-orthogonality complicates the application of conventional quantum algorithms, such as quantum phase estimation (QPE), for energy calculations. To overcome this challenge, we introduce the unitary PAW (UPAW) method, which ensures the unitarity of the transformation and as a result preserves the orthonormality of wavefunctions.
Our formulation of the many-body UPAW is general and can be applied in either first or second quantization. 

%
%
One of the advantages of PAW is the reduction of the two-body rank, which directly affects on the compactness of the linear combination of unitaries (LCU) decomposition - a crucial representation used in many quantum algorithms, including qubitization-based \cite{low_hamiltonian_2019,poulin_quantum_2018,berry_improved_2018} quantum phase estimation (QPE) \cite{Kitaev1995,NielsenChuang} considered in this study. Prior work has developed such LCU decompositions and corresponding block encodings for molecular systems~\cite{berry_qubitization_2019,von_burg_quantum_2021,lee_even_2021} and, more recently, for periodic solids using Gaussian basis sets~\cite{ivanov_quantum_2023,rubin_fault-tolerant_2023}. Here, we extend these techniques to construct an explicit LCU and block encoding for the UPAW Hamiltonian in second quantization using KS orbitals and a plane-wave basis set. Specifically, we express the Hamiltonian as an LCU
\begin{equation}\label{eq:lcu} \hat H = \sum_{l=0}^{L-1} \omega_l \hat W_l, \end{equation} where $\omega_l$ are real coefficients and $\hat W_l$ are unitary operators. For use in qubitization-based QPE, a quantum circuit construction embeds the Hamiltonian into a larger unitary matrix $\hat V$, a so called block encoding:
\begin{equation} 
\hat V = 
    \begin{pmatrix} \hat H / \lambda & * \\ * & * \
    \end{pmatrix}, 
\end{equation}
where the subnormalization factor $\lambda$ is typically given by the one-norm $\lambda = \sum_{l=0}^{L-1} |\omega_l|$. The efficiency of qubitized QPE is determined by $\lambda$ and $\Gamma$, the amount of information needed to specify the LCU decomposition~\eqref{eq:lcu}, i.e.~the total number of bits of all coefficients. Using a QROAM-based data-loading scheme~\cite{low2018trading,berry_qubitization_2019}, the Toffoli gate complexity scales as $O(\sqrt{\Gamma} \lambda / \epsilon_{\rm QPE})$, and the required number of qubits scales as $O(\sqrt{\Gamma})$, where $\epsilon_{\rm QPE}$ is the error tolerance. Since these parameters depend on the choice of basis, an important question we address is how they scale for physical systems when using the UPAW method.

Furthermore, in this work we leverage a popular strategy developed in classical computation called down-sampling~\cite{ohnishi_logarithm_2010,gruneis_natural_2011}. Using down-sampling, the energy of the system can be estimated using energy differences obtained with a smaller number of orbitals. This results in lower computational resources to estimate the energy at the converged basis set limit. Naturally, the question arises if such a technique would provide feasible quantum resources for ground-state energy estimation. We will explore this question in Sec.~\ref{sec:numerical results}.

To summarise, the results of this work are: (i) We introduce the unitary PAW method, a generalization of PAW for many-body wavefunctions; (ii) We present an LCU decomposition of the Hamiltonian in second quantization using UPAW and plane waves and use it in qubitization-based QPE. Our asymptotic scalings are similar to those from double-factorization based on a Cholesky decomposition~\cite{von_burg_quantum_2021} when the system grows towards the thermodynamic limit; (iii) We carry out resource estimation for diamond using the down-sampling technique for calculating the energy within chemical accuracy with respect to the full basis set limit; (iv) Finally, we also consider the nitrogen-vacancy centre in diamond -- a challenging system for classical algorithms, and provide the resource estimates for this system too.

The paper is organized as follows. First, in Sec.~\ref{sec:paw_method}, we briefly review the PAW method before introducing the many-body PAW transformation in Sec.~\ref{sec:mb-paw_method} and a unitary version of the method, UPAW, in Sec.~\ref{sec:upaw_method}. We then turn to its implementation on a quantum computer, presenting  an LCU decomposition of the Hamiltonian using UPAW and plane-waves in Sec.~\ref{sec:hamiltonain}, which is essential for the use in qubitized QPE. In Sec.~\ref{sec:block-encoding}, we outline the application of quantum circuits from the double-factorization method to our setting. In Sec.~\ref{sec:asymptotics}, we numerically determine asymptotic scalings of the quantum computational time and space cost of the algorithm. Further in Sec.~\ref{sec:quantum resource estimates}, we show quantum resource estimates for periodic solids which includes crystalline diamond and the nitrogen-vacancy defect center in it. In Sec.~\ref{sec:discussion_and_conclusion}, discussion and conclusion are presented.

\section{Theory and Implementation}
\subsection{Review of Projector Augmented-Wave (PAW) method}
\label{sec:paw_method}

\begin{figure}
    \centering
    \includegraphics[width=0.9\linewidth]{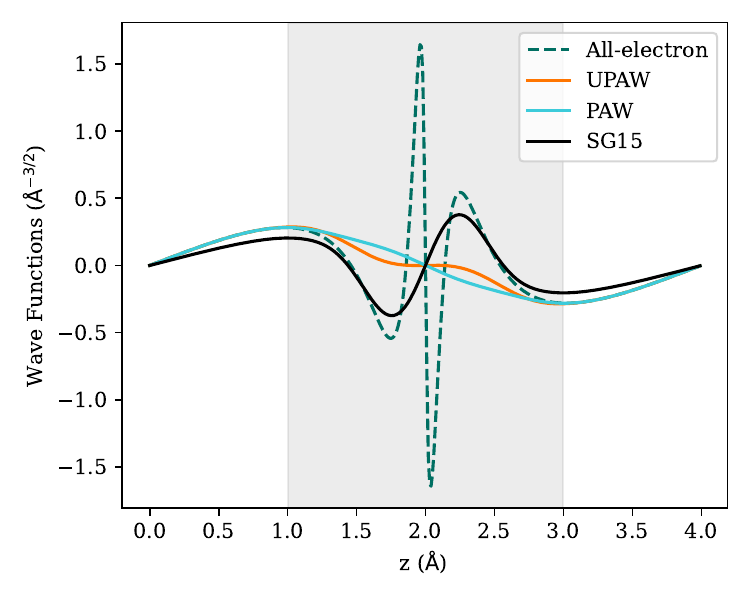}
    \caption{{\bf All-electron and pseudo orbitals obtained using PAW~\cite{blochl_projector_1994}, UPAW [this work] and norm-conserving Vanderbilt pseudopotential (SG15)~\cite{hamann_optimized_2013}.} Orbitals (lowest unoccupied orbitals + 1) obtained from calculations of Ni atom using 800 eV kinetic energy cutoff. The grey interval indicates an area outside of which the all-electron and (U)PAW orbitals are the same. In 3D, this interval would instead be a ball.}
    \label{fig:pseudo-orbitals}
\end{figure}

Before introducing the many-body version of PAW, we summarize the original PAW approach~\cite{blochl_projector_1994} applied to KS-DFT. The problem that PAW addresses is the fact that the wavefunction shows different behaviour in different region of space. The wavefunction exhibits rapid oscillations near the nuclei while it is smooth elsewhere. A key feature of PAW is that it establishes a linear transformation $\hat{\mathcal{T}}$ between a pseudo Hilbert space~\cite{blochl_projector_1994} and the Hilbert space, mapping smooth pseudo wavefunctions (pseudo orbitals) $\{\ket{\widetilde{\psi}}\}$ to full wavefunctions (orbitals) $\{\ket{\psi}\}$:
\begin{equation}
\hat{\mathcal{T}} \ket{\widetilde{\psi}} = \ket{\psi}.
\end{equation}
This linear transformation should be defined in such a way that pseudo orbitals do not have cusps and rapid oscillations around the nuclei and coincide with full orbitals in other region of space.
%
To achieve this, the system is divided into atom-centered augmentation spheres within which $\ket{\psi}$ is expanded in terms of localized (atom-like) partial waves $\ket{\phi_{j}^{a}}$, where $j$ identifies one such function belonging to the atom labelled by $a$. Similarly, the pseudo orbitals $\ket{\widetilde{\psi}}$ are also expanded using smooth pseudo partial waves denoted as $\ket{\widetilde{\phi}_{j}^{a}}$. The functions $\ket{\phi_{j}^{a}}$ and $\ket{\widetilde{\phi}_{j}^{a}}$ are identical outside the augmentation sphere $\mathbb{B}_a$ around atom $a$, and they are related by the transformation $\hat{\mathcal{T}}$ in the same way as the orbitals $\ket{\psi}$ and $\ket{\widetilde{\psi}}$. As a result, from the described atomic-centre expansion of orbitals, the PAW operator, $\Tau$, can be given as follows~\cite{blochl_projector_1994}:
\begin{equation}\label{eq:paw_transform_original}
    \Tau = I +  \sum_{a=1}^{N_A}\sum_{j=1}^{n_a} \left(\ket{\phi_{j}^{a}} - \ket{\widetilde{\phi}_{j}^{a}}\right) \bra{\tilde{p}_{j}^{a}}
\end{equation}
%
Here, $N_a$ is the total number of atoms in the system, and  $n_a$ is the total number of projector functions, $\{\ket{\tilde{p}_{j}^{a}}\}$, localized on an atom $a$, which can be constructed from a set of linearly independent functions and must be biorthogonal to pseudo partial waves, $\ket{\widetilde{\phi}_{j}^{a}}$~\cite{blochl_projector_1994}. 
 To simplify our notation, the combined index $k = (a, j)$ will be introduced as well as the functions:
\begin{align}
    & \ket{\chi_k} = \ket{\phi_k} - \ket{\widetilde{\phi}_k}.
\end{align}
Both $\ket{\chi_k}$ and $\ket{\tilde{p}_k}$ are localized at the nucleus $a$ and vanish outside the augmentation sphere.
As an example, we show the results of the PAW transformation in Fig.~\ref{fig:pseudo-orbitals}. We note that the operator $\Tau$ is not generally unitary. This can be confirmed by calculating the overlap matrix:
\begin{equation}\label{eq:overlap_one_electron}
    \hat O := \hat{\mathcal{T}}^{\dagger} \hat{\mathcal{T}} = I +  \Delta \hat O,
\end{equation}
where 
\begin{align}
    \Delta \hat O = \sum_{a} \sum_{i, j} \ket{\tilde{p}_{i}^{a}} O_{ij}^a \bra{\tilde{p}_{j}^{a}}, \\
     O_{ij}^a = \braket{\phi_{i}^a|\phi_{j}^a} - \braket{\tilde{\phi}_{i}^a|\tilde{\phi}_{j}^a}, 
\end{align}
Because $\Tau$ is not unitary, this means that pseduo-orbitals, and as a result the eigenfunctions of the pseudo Hamiltonian, are not orthogonal. This complicates the use of PAW for some quantum algorithm.

\subsection{Many-body Projector Augmented-Wave (many-body PAW) method}\label{sec:mb-paw_method}

Next, we show how to generalize the PAW transformation to a many-body wavefunction, $\Psi$. In order to do this, we follow a procedure, similar to the configuration interaction expansion~\cite[p.251]{mayer_simple_2003}, in which the PAW transformation is applied recursively coordinate by coordinate:
\begin{multline}
    \Psi_{j-1}(\tx_1, \tx_2, \dots, \tx_N)  = 
    \Psi_{j}(\tx_1, \tx_2, \dots, \tx_N) + {} \\
    \sum_{k_j} \chi_{k_j}(\tr_j) \int d\tr_j' \, \tilde{p}_{k_j} (\tr_j') \Psi_{j}(\tx_1, \dots, \tx_j', \dots, \tx_N), \quad j=1 \dots N,
\end{multline}
where $N$ is the number of electrons, $\tx=(\tr, \sigma)$ is a combined coordinate variable consisting of the spatial ($\mathbf{r}$) and spin ($\mathbf{\sigma}$) degrees of freedom, $\Psi_{0}$ is the original wavefunction and $\Psi_{N}$ is a smooth wavefunction that does not have cusps and rapid oscillations around the nuclei,
\begin{align}
    \Psi_{0}(\tx_1, \tx_2, \dots, \tx_N) := \Psi(\tx_1, \tx_2, \dots, \tx_N), \\
    \Psi_{N}(\tx_1, \tx_2, \dots, \tx_N) := \tilde{\Psi}(\tx_1, \tx_2, \dots, \tx_N).
\end{align}
Using this recursive relation, one obtains the many-body PAW transformation:
\begin{multline}    
    \hat{\mathcal{T}}_{MB} = 
    I + \sum_{i_1} 
    \sum_{k_1}
    \ket{\chi_{k_1}(i_1)}\bra{\tilde{p}_{k_1}(i_1)} + {} \\ 
    \sum_{i_1 < i_2}
    \sum_{k_1 < k_2}
    \ket{\chi_{k_1}(i_1) \chi_{k_2}(i_2) }
    \bra{\tilde{p}_{k_1}(i_1) \tilde{p}_{k_2}(i_2)} + \dots  + {} \\ 
    \sum_{k_1 < \dots < k_N} 
    \ket{\chi_{k_1}(1) \dots \chi_{k_N}(N)}  \bra{\tilde{p}_{k_1}(1) \dots \tilde{p}_{k_N}(N)},
\end{multline}
where $\ket{\chi_{k_1}(1) \dots \chi_{k_N}(N)}$, $\ket{\tilde{p}_{k_1}(1) \dots \tilde{p}_{k_N}(N)}$ are the Slater determinants, and the PAW-transformed Schrödinger equation for the pseudo many-body wavefunction can be then written as follows:
\begin{equation}
 \hat{\tilde{H}} \ket{\tilde{\Psi}} = E \hat S   \ket{\tilde{\Psi}},
\end{equation}
with 
\begin{align}
     \hat{\tilde{H}} & = \hat{\mathcal{T}}_{MB}^{\dagger} \hat{H} \hat{\mathcal{T}}_{MB},  \\
     \hat S & = \hat{\mathcal{T}}_{MB}^{\dagger} \hat{\mathcal{T}}_{MB}.
\end{align}
Here $E$ represents the energy of the all-electron system, and $H$ is the Hamiltonian of interest.
While the wavefunction becomes smooth around nuclei, the PAW transformation brings additional complications. First, as already was mentioned above, the eigenstates are not orthonormal anymore:
\begin{equation}
    \braket{\tilde{\Psi}_i|\tilde{\Psi}_j} \neq \delta_{ij}.
\end{equation}
As a result, one has to solve a generalized eigenvalue problem.
Secondly, the PAW transformation generally leads to an $N$-body interaction term unless one implements the inverse PAW transformation, $\hat{\mathcal{T}}_{MB}^{-1}$. This complicates the use of orthonormal basis sets such as plane waves, while the use of non-orthogonal basis functions complicates the mathematical description of many-body techniques.
In order to overcome these shortcomings we develop the \textit{unitary} PAW method (UPAW) which eliminates the complications of a conventional PAW method.

\subsection{Unitary Projector Augmented-Wave (UPAW) method}
\label{sec:upaw_method}

In our unitary variant, the general structure of the PAW method will be largely unchanged. Modifications are introduced at the level of constructing pseudo partial waves from Eq.~\ref{eq:paw_transform_original}. We would like to make $\hat{\mathcal{T}}$ a unitary operator, $\hat{\mathcal{T}}^{\dagger} = \hat{\mathcal{T}}^{-1}$. This means that the one-electron overlap operator from Eq.~\ref{eq:overlap_one_electron}
becomes the identity. This requires us to choose pseudo partial waves $\tilde{\phi}_i^a$ in such a way that $O_{ij}^a$ is zero:
\begin{equation}\label{eq:upaw_condition}\braket{\tilde{\phi}_{i}^a|\tilde{\phi}_{j}^a}  = \braket{\phi_{i}^a|\phi_{j}^a}. 
\end{equation}
 The partial and pseudo partial waves are atomic-like orbitals, $\phi_i(\tr) = r^l R_{nlp}(r) Y_{lm}(\hat \tr)$, where $n$ is the principle quantum number, $l$ is the angular momentum, $m$ is magnetic quantum number and $p$ is an additional index enumerating the number of partial waves per angular momentum value. While partial waves are solutions of the atomic Schrodinger equation, the pseudo partial waves $\tilde{\phi}_i^a$ are constructed to satisfy the properties:
\begin{enumerate}
    \item The radial part of $\tilde{\phi}_i^a(\tr)$, $\tilde{R}_{nlp}(r)$, is a smooth function for $r \in \mathbb{B}_a$.
    \item $\tilde{R}_{nlp}(r) = R_{nlp}(r)$ for $r \in \mathbb{\overline{B}}_a$.
    \item At the boundary, we require  $\partial^{i} \tilde{R}_{nlp}(r_a) = \partial^{i} R_{nlp}(r_a)$, $i \in \{0, 1\dots, P-1\}$ up to a judiciously chosen order $P$. 
    \item $O_{ij}^a = 0$ for any $i, j, a$.
\end{enumerate}
Conditions 1--3 are conventional conditions used to construct pseudo partial waves while the orthonormality condition 4 is a new additional constraint that makes the operators unitary.
We use a polynomial approximation to represent the radial part of the pseudo partial waves inside the augmentation spheres:
\begin{equation}
    \tilde{R}(r) = \sum_{p=0}^{P + M - 1} (r^2)^{P + M -1 - p} c_p, 
\end{equation}
where we allocate an extra $M$ coefficients to satisfy the orthonormality constraints~\ref{eq:upaw_condition}. 
The coefficients $c_p$ are determined by non-linear optimization until the constraints 1--4 are satisfied. In order to improve the smoothness of the pseudo partial waves, for some atoms the optimization procedure also attempts to remove high Fourier components related to pseudo partial waves. Namely, let $b_i^{a}(\tG)$ be the Fourier transform of $r \tilde{\phi}_i^a(\tr)$, then we set $\sum_{|\tG| > G_{max}} |\tG|^2 |b_i^{a}(\tG)|^2=0$, where $G_{max}$ is a parameter. An example pseudo partial wave in UPAW is shown in Fig.~\ref{fig:pseudo-orbitals}, and we will discuss the quality of the produced UPAW setups for some chemical elements in Sec.~\ref{sec:numerical results}. With the construction of the pseudo partial waves now completed, we continue with the formulation of the many-body problem.

For the unitary PAW transform, the pseudo many-body Schr\"odinger equation becomes 
\begin{equation}
    \hat{\tilde{H}} \tilde{\Psi} = E \tilde{\Psi},
\end{equation}
where
\begin{equation}
\label{eq:upaw_hamiltonian} 
    \hat{\tilde{H}} = C + \sum_{j=1}^N \hat{\mathcal{T}}_j^{\dagger} \hat h_j \hat{\mathcal{T}}_j + \frac{1}{2}\sum_{i\neq j}^N\hat{\mathcal{T}}^{\dagger}_i \hat{\mathcal{T}}^{\dagger}_j  \hat g_{ij} \hat{\mathcal{T}}_i \hat{\mathcal{T}}_j    
    = C + \sum_{j=1}^N  \hat{\tilde{h}}_j + \frac{1}{2}\sum_{i\neq j}^N \hat{\tilde{g}}_{ij}.    
\end{equation}
where $\hat h$ and $\hat g$ are the conventional one- and two-body operators. For practical applications, one must evaluate the matrix elements of the Hamiltonian in either first or second quantization:
\begin{equation}\label{eq:ftq_ham}    
    \hat{\tilde{H}} = 
    C +   
    \sum_{i}^{N} 
    \sum_{pq}^{N_b} 
        h_{p q} 
    \sum_{\sigma}(\ket{\tilde{p} \sigma}\bra{\tilde{q}\sigma})_i  
    + 
    \frac{1}{2}
    \sum_{i\neq j}^{N}
    \sum_{pqrs}^{N_b}
        \kappa_{q p r s}
    \sum_{\sigma, \tau}
    (\ket{\tilde{p}\sigma}\bra{\tilde{q}\sigma})_i
    (\ket{\tilde{r}\tau}\bra{\tilde{s}\tau})_j,
\end{equation}    
and
\begin{equation} \label{eq: Ham-Ferm-Chem}
    \hat{\tilde{H}} = \sum_{pq}^{N_b}\left(h_{pq} - \frac{1}{2} \sum_{r}^{N_b} \ka_{rprq}\right) \hat{E}_{pq} + \frac{1}{2} \sum_{pqrs}^{N_b} \ka_{qprs} \hat{E}_{pq}\hat{E}_{rs},
\end{equation}
respectively. In the expression above, we used a spin-restricted approach and an orthonormal smooth basis set $
\{\ket{\tilde{p} \sigma}\}$, where each function is labelled by a number $\tilde{p}$ and spin index $\sigma$; $N_b$ is the number of basis functions discretising the Hamiltonian. Spin-summed excitation operators (excitons) $\hat{E}_{pq}$ are defined as
\begin{equation}    
\hat{E}_{pq} = \sum_{\sigma=0,1}\hat{a}_{p\sigma}^\dagger \hat{a}_{q\sigma}. 
\end{equation}
The one- and two-body matrix elements calculated from the PAW-transformed one- and two-body operators are: 
\begin{equation}
     h_{p q} = \braket{\tilde{p}| \Tau^{\dagger} \hat h \Tau|\tilde{q}},
\end{equation}
\begin{equation}~\label{eq: kappa_pqrs_all_electron}
    \kappa_{qprs} = \bra{\tilde{p}}\bra{\tilde{r}} \Tau^{\dagger} \otimes \Tau^{\dagger} \hat g \Tau \otimes \Tau \ket{\tilde{q}}\ket{\tilde{s}}.
\end{equation}
The operators $\ket{\tilde{p} \sigma}\bra{\tilde{q}\sigma}$, $\adps \aqs$ in Eqs.~\ref{eq:ftq_ham} and \ref{eq: Ham-Ferm-Chem} act directly on the pseudo many-body wavefunctions that require a significantly smaller number of basis functions, especially if plane wave basis sets are used. 

\subsection{Linear Combination of Unitaries Decomposition}\label{sec:hamiltonain}

In this work we focus on quantum computation in second quantisation. 
In order to implement the Hamiltonian~\ref{eq: Ham-Ferm-Chem} with low-cost quantum circuit, it is necessary to factorize the Hamiltonian so as to reduce the amount of information needed for the block encoding and to have the smallest possible subnormalization factor. First, we notice that the two-body term~\ref{eq: kappa_pqrs_all_electron} in the Hamiltonian can be expanded into a soft (pseudo) contribution and atomic-centered PAW corrections~\cite{rostgaard_projector_2009}:
\begin{equation}\label{eq:two_doby_decomposition}
    \ka_{pqrs} =(\tilde{\rho}_{pq}|\tilde{\rho}_{rs}) +  \sum_{a}^{N_A}
    \sum_{i_1i_2i_3i_4}^{n_{a}} C^{a}_{i_1 i_2 i_3 i_4} D_{pq,i_1i_2}^{a*} D_{rs,i_3i_4}^{a},
\end{equation}
respectively, where we introduced the notation for Coulumb matrix elements:
\begin{equation}
    (f|g) = \IINTr{} \frac{f^*(\tr') g(\tr)}{|\tr' - \tr|}.
\end{equation}
For the definition of $\tilde{\rho}_{pq}$ and other notations we refer the reader to Appendix~\ref{sec: matrix elements}. We factorize the first term in Eq.~\ref{eq:two_doby_decomposition} with the plane wave expansion of the Coulomb kernel
\begin{equation}
    \frac{1}{|\tr|} = \sum_{\tG}^{N_{\rm pw}} e^{i\tG\tr} v({\tG}).
\end{equation}
In this expansion, $N_{\rm pw}$ is the total number of plane waves. We do not omit the divergent $\tG=0$ component and instead use Wigner-Seitz regularization~\cite{sundararaman_regularization_2013}. For the PAW term, we use eigendecompositions of the tensor $C^{a}_{i_1i_2i_3i_4}$ and the atomic orbital-pair density matrices, $D_{pq,i_1i_2}^{a}$. 
Due to the length and complexity of the derivation, which involves tedious mathematical steps, we present only the final result here and refer the reader to Appendix~\ref{sec: matrix elements} for the full parametrization of the Hamiltonian and a detailed derivation with explanation of all notation. The  Hamiltonian can then be written as follows:

\begin{multline}  \label{eq:full_hamiltonian}  
    \hat{\tilde{H}} =  
    \sum_{p,q}^{N_b} \left(h_{pq} - \frac{1}{2}\sum_{r}\kappa_{rprq}\right)\hat{E}_{pq}
    +
    2\sum_{\tG \geq 0}^{N_{\rm pw}/2} \sum_{j=1,2} v'(\tG) \hat U_{j}(\tG) \left(\sum_{p}^{N_{b}} f_{p,j}(\tG) \hat{E}_{pp}\right)^2 \hat U^{\dagger}_{j}(\tG)
    + \\ 
    \sum_{a}^{N_A} \sum_{i_1\leq i_2}^{n_a} {\rm sign}(\epsilon_{i_1i_2}^a) \hat U_{i_1i_2}^a\left( \sqrt{|\epsilon_{i_1i_2}^a|} \sum_{p}^{N_b} f_{p, i_1i_2}^a \hat{E}_{pp} \right)^2 \hat U_{i_1i_2}^{a\dagger}
\end{multline}
where $\hat U_j$ are unitary operators and $f_{p,j}(\tG)$ are the eigenvalues of the soft reciprocal orbital-pair density matrix which give the most dominant contributions to the subnormalization factor (see details in Appendix~\ref{sec: matrix elements}). When the Hamiltonian~\ref{eq:full_hamiltonian} is mapped onto qubits, an additional contribution appears in the one-body term~\cite{von_burg_quantum_2021,lee_even_2021}
\begin{equation}\label{eq:modified_one_body_qubit}
    h'_{pq} = h_{pq} + \frac{1}{2}\sum_{r} (2\kappa_{rrpq} - \kappa_{rprq}),
\end{equation}
and the subnormalization factor in the block encoding becomes
\begin{equation} \label{eq:subnormalisation}
    \lambda = 
    \sum_{p}^{N_b} |\epsilon_p| +
    \frac{1}{4} \sum_{j=1,2} \sum_{pq}^{N_b} \xi^{(j)}_{pq} + \frac{1}{2} \sum_{a}^{N_A}\sum_{i_1\leq i_2}^{n_a} |\epsilon_{i_1i_2}^{a}| \left( \sum_{p} |f_{p, i_1i_2}^{a}| \right)^2,
\end{equation}
where $\epsilon_p$ is the $p$\textsuperscript{th} eigenvalue of $h'$~\cite{von_burg_quantum_2021},
and 
\begin{equation}    
    \xi_{pq}^{(j)} =
    \frac{8\pi }{V} \sum_{\tG>0} \frac{1}{\tG^2}|f_{p, j}(\tG)| |f_{q, j}(\tG)|
    = 
    \IINTr{V}\frac{f_{p,j}(\tr) f_{q,j}(\tr')}{| \tr - \tr'|} = 
    ( f_{p,j} | f_{q,j}).
\end{equation}
The additional term in equation~\ref{eq:modified_one_body_qubit} arises because $\hat E_{pp}$ can be represented as a sum of $Z$ and identity operators in a qubit representation. This combination leads to an effective one-body contribution after the sum of $\hat E_{pp}$ is squared in equation~\ref{eq:full_hamiltonian}. 
The factorization~\ref{eq:full_hamiltonian} is of double factorization form~\cite{motta_low_2021, von_burg_quantum_2021} apart from the fact that there is an additional sign in the PAW contribution, ${\rm sign}(\epsilon_{i_1i_2}^a)$. This is because $C_{i_1i_2i_3i_4}^{a}$  is not positive definite and the standard Cholesky decomposition could not be used. Instead, we have used a conventional eigendecomposition. In the next Section~\ref{sec:block-encoding}, we outline the modifications to a quantum circuit from Ref.\citenum{lee_even_2021} in order to take the two-body PAW term into account.

\subsection{Block Encoding}\label{sec:block-encoding}

\begin{figure}
 \centering 
	   \includegraphics[width=\textwidth]{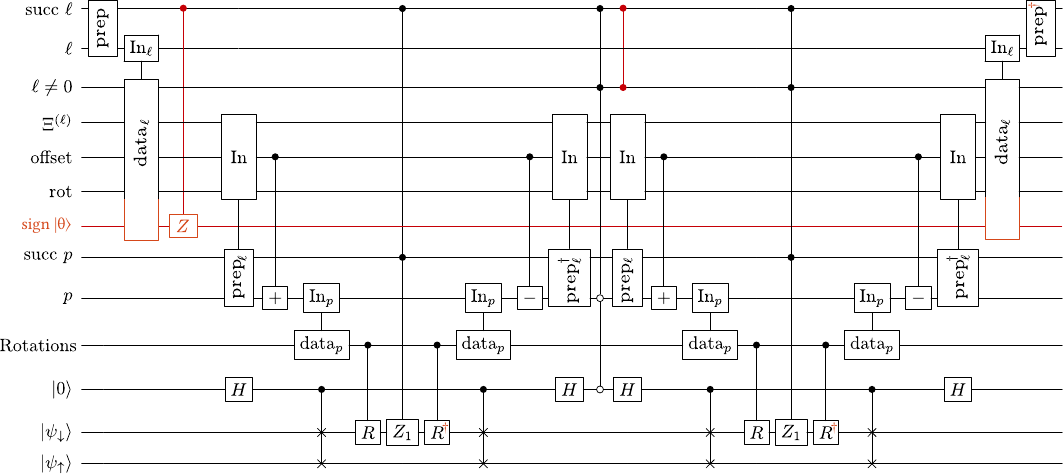}
 \caption{\textbf{Circuit for the block encoding of the double factorised UPAW Hamiltonian.}  The image is a modified version from Ref.\citenum{lee_even_2021} figure 16 (available under the terms of the Creative Commons Attribution 4.0 International license Copyright 2021). Compared to that reference, the modifications in red enable the signs required by the UPAW contribution to the interaction term.}
 \label{fig:block_encoding}
\end{figure}

Our block encoding of the UPAW Hamiltonian closely follows double factorization \cite{von_burg_quantum_2021} as presented in \cite[Appendix C]{lee_even_2021}. Fig.~\ref{fig:block_encoding} shows the block encoding circuit of \cite{lee_even_2021} along with our modifications in red. First, the one-body term and factorised terms in the Hamiltonian (\ref{eq:full_hamiltonian}) are indexed by $\ell= 0,1,...,L$. While $\ell=0$ flags the one-body terms, positive values of $\ell$ flag the various factorised two-body terms indexed by $\tG, j, a, i_1, i_2$ in \eqref{eq:full_hamiltonian}. Counting those terms results in 
\begin{equation}L=N_{\rm pw}+\sum_{a}^{N_A} n_a(n_a+1)/2.\label{eq:L}\end{equation}
The key difference of the UPAW Hamiltonian compared to the prior work is the coefficients ${\rm sign}(\epsilon_{i_1,i_2}^a)$ in the UPAW contribution, Eq.~\ref{eq:paw_interaction_term}, to the interaction term. These signs cannot be absorbed into a square with real coefficients and therefore must be treated separately. This can be achieved \cite{von_burg_quantum_2021} in the circuit as shown in Fig.~\ref{fig:block_encoding}: The signs are loaded into an additional ancilla $\ket{\theta}$ (with $\ket{0}$ indicating $+$ and $\ket{1}$ indicating $-$) during the data loading indexed by the $\ell$ register. We use ${\rm sign}(\epsilon_{i_1,i_2}^a)$ for the UPAW contribution to the interaction term and $+1$ for all other terms. The phase corresponding to the sign is then implemented by acting on this ancilla with a $Z$ gate, conditioned on the success of the state preparation over $\ell$.
We also add the complex conjugation to the uncomputation of the $R$ rotations that were missing in the original circuit.
The square from the factorization is implemented by performing the inner block encoding (of the base of the square) twice in the case of two-body terms $\ell \neq 0$, with an intermediate reflection, to which we have added a CZ to correctly recover the second Chebyshev polynomial, see Appendix~\ref{app: square}.

This extension of the double factorization circuit to implement UPAW only incurs a negligible increase in fault-tolerant resources: Firstly, the output register of the unary iteration over $\ell$ needs one additional qubit to store the sign of the interaction terms. The increased register size also leads to a slight additive increase in Toffoli cost for implementing the QROAM for the data loading indexed on $\ell$.
As the CZ implementing the sign is a Clifford gate, it does not contribute to the fault-tolerant resource cost (see Appendix~\ref{app:error correction} describing the error correction scheme).
Following the convention of \cite[Appendix C]{lee_even_2021}, where costing of the double factorisation algorithm is re-presented, this amounts to increasing \cite[Eq.~C29]{lee_even_2021} for $b_o$ by one. In the OpenFermion \cite{mcclean2020openfermion} package's implementation of the double factorisation costing, this modification can be readily achieved by increasing the $b_o$ variable by one.

\section{Results}\label{sec:numerical results}

\subsection{Space and Time Complexity of the Quantum Algorithm}\label{sec:asymptotics}

In this section, we present the asymptotic complexity of the algorithm in the worst-case scenario, when only terms in the Hamiltonian with absolute value less then $10^{-10}$ are neglected. Such a small truncation provides results well within chemical accuracy. The total Toffoli count of qubitized QPE is $O(\sqrt{\Gamma}\lambda/\epsilon_{\rm QPE})$, with $O(\sqrt{\Gamma})$ qubits~\cite{berry_qubitization_2019,low2018trading}.
The subnormalisation (one-norm) $\lambda$ and target QPE error $\epsilon_{\rm QPE}$ give the number of iterations of the block encoding required. The cost of the block encoding is dominated by a QROAM circuit loading data of total size $\Gamma$. Here, the main contribution to the data that must be loaded to implement the Hamiltonian is the $L$ rotations (see Eq.~\ref{eq:L}) for the two-body terms of dimensions $N_b\times N_b$. Each rotation angle is specified with $\beth$ bits, resulting in total data size $    \Gamma=LN_b^2\beth$, like in double-factorization \cite{von_burg_quantum_2021}. Specifically, for our $LN_b$ data items of size $N_b\beth$ each (rotations are loaded column-by-column), the QROAM allows a space-time tradeoff
\begin{equation}
    \text{QROAM Toffolis:\ } \left\lceil\frac{LN_b}{k_r}\right\rceil + N_b\beth (k_r-1) \quad \text{QROAM qubits:\ }N_b\beth k_r + \left\lceil\log\frac{LN_b}{k_r}\right\rceil
\end{equation}
with a tunable parameter\footnote{For simplicity, we assume $k_r$ to be a power of 2.} $k_r$. When it is chosen to minimise Toffoli count, we recover Toffoli and qubit cost $O(\sqrt{\Gamma})$ for the block encoding.

In practice, we have the scaling $\Gamma=O(N_{\rm pw}N_b^2)$, because $N_{\rm pw} \gg N_A$, and $\beth$ can be ommited as it only weakly depends on the size of the system and in this work we kept it constant ($\beth=20$) as will be discussed in Section~\ref{sec:quantum resource estimates}.
For the subnormalisation, one can expect that $\lambda=O(N_b^2),$ by using the assumptions $\xi^{(j)}_{pq}=O(1)$ in Eq.~\ref{eq:subnormalisation} and that the last term in the equation is small (as verified numerically).
We will further analyse the complexity of the algorithm in two regimes, the continuum limit and the thermodynamic limit.

The first regime we consider is the continuum limit, in which system size $N_A$ is fixed, and the number of bands $N_b\to\infty$ increases. While at first glance this requires $N_{\rm pw}=O(N_b)$, in practical calculations, $N_{{\rm pw}}$ is always significantly larger than $N_{b}$, especially for molecules due to the large simulation box needed to reduce the interaction between periodic images. For any practical calculation when natural orbitals are employed (see Appendix~\ref{sec: mp2no}), the correlation energy converges for $N_{b} \ll N_{\rm pw}$~\cite{gruneis_natural_2011} such that $N_{\rm pw}$ can be taken as a large, fixed constant. Consequently, we expect $\Gamma =O(N_b^2)$ and total Toffoli complexity $O(N_b\lambda)$ and space complexity $O(N_b)$ for the convergence of correlation energy w.r.t~the continuum limit. This indicates that the qubit complexity would scale only linearly in the number of orbitals, until it reaches the plane-wave basis set size.

\begin{figure}  
 \centering
	   \includegraphics[width=1.0\textwidth]{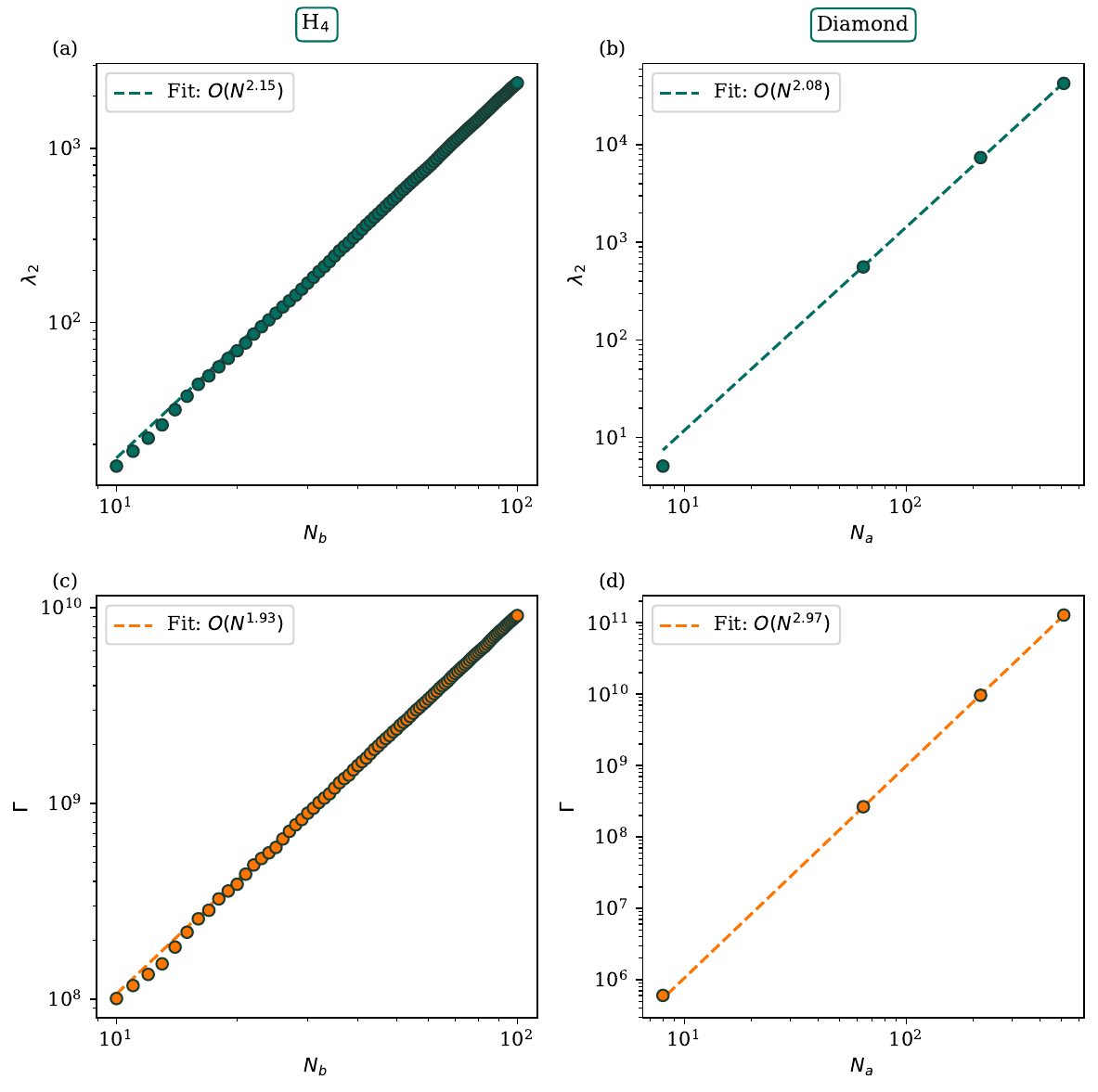}
 \caption{{\bf Scaling of different parameters that determine the efficiency of the quantum algorithm.} (a, b) The subnormalization factor $\lambda_2$ in the block-encoding due to the two-body term. (c, d) The amount of information needed to specify the Hamiltonian's linear-combination-of-unitaries decomposition, $\Gamma$. Only parameters with absolute value larger than $10^{-10}$ are counted as being non-zero. For the H$_4$ example (a, c), the number of atoms and plane waves is fixed and only the number of orbitals is changed while for Diamond (b, d) the number of atoms is changed which in turn also changes the total number of orbitals and plane waves.}
 \label{fig:scalingplot}
\end{figure}

We verify this scaling numerically with a test system of four hydrogen atoms arranged in a square with length 2 Bohr, see Fig.~\ref{fig:scalingplot}(a,c).
Since the two-body term is the dominant contribution in the asymptotic regime, we considered only its contribution to the one-norm:
\begin{equation}\label{eq:two_norm_two_body}    
    \lambda_2 = 
    \frac{1}{4} \sum_{j=1,2} \sum_{pq}^{N_b} \xi^{(j)}_{pq} + \frac{1}{2} \sum_{a}^{N_A}\sum_{i_1\leq i_2}^{n_a} |\epsilon_{i_1i_2}^{a}| \left( \sum_{p} |f_{p, i_1i_2}^{a}| \right)^2.
\end{equation}
For the molecular hydrogen, we used a $10\times10\times10$ {\AA}$^3$ box and the kinetic energy cutoff was set to 800 eV. As in Ref.\citenum{lee_even_2021}, we considered varying numbers of orbitals from 10 to 100. The results in Fig.~\ref{fig:scalingplot}(c) show that $\Gamma$ scales almost quadratically $\Gamma=O(N_b^{1.93})$, and Fig.~\ref{fig:scalingplot}(a) shows that the subnormalization scales as $\lambda=O(N_b^{2.15})$, both close to the theoretical scalings discussed above.

The second regime we consider is the thermodynamic limit, in which supercell size and number of atoms $N_A\to\infty$ increases, and the number of bands per atom $n_b = N_b/N_A$ is fixed. The number of plane waves must also increase linearly with the number of atoms, $N_{\rm pw} = O(N_A)$. Consequently, we expect $\Gamma = O(N_A^3)$ and space complexity $O(N_A^{1.5})$. Again, we calculate this scaling numerically. To that end, we consider crystalline diamond with supercells of 2--128 atoms with a fixed number of orbitals per atoms (4 natural orbitals per atom). We use a kinetic energy cutoff of 600 eV. The results are presented in Fig.~\ref{fig:scalingplot}(b, d). We see that the one-norm also scales almost quadratically $O(N_A^{2.08})$, and $\Gamma=O(N_A^{2.97})$, close to the theoretical scalings discussed above. The asymptotic results in this regime are summarized in Table~\ref{tab:scaling_comparison}, where for comparison we also provide asymptotic scalings of other factorization techniques.

\begin{table}
    \centering
    \begin{tabular}{c|c|c}
    \hline
     Factorization &  Qubit Complexity & Toffoli Complexity \\
    \hline
     Sparse~\cite{berry_qubitization_2019} &  $N_A + \sqrt{S}$ &  $ (N_A + \sqrt{S}) \lambda_{\rm Sparse} / \epsilon$ \\
     SF~\cite{berry_qubitization_2019} &  $N_A^{3/2}$ &  $N_A^{3/2} \lambda_{\rm SF} / \epsilon$ \\
     DF~\cite{von_burg_quantum_2021} &  $N_A\sqrt{\Xi}$ &  $N_A\sqrt{\Xi} \lambda_{\rm DF} / \epsilon$ \\
     THC~\cite{lee_even_2021} &  $N_A$ &  $N_A \lambda_{\rm THC} / \epsilon$ \\
     PW-PAW [This work] & $N_b \sqrt{N_{\rm pw}} \propto N_A^{3/2}$ & $ N_A^{3/2} \lambda_{2} / \epsilon$ \\
    \end{tabular}
    \caption{\textbf{Asymptotic complexities of different factorization algorithms for 3D periodic systems w.r.t.~the number of atoms (system size) $N_A$, and a fixed number of orbitals per atom.} $N_b$ is the number of bands (orbitals) and $N_{\rm pw}$ is the number of plane waves. SF -- Single Factorization, DF -- Double Factorization, THC -- Tensor Hypercontraction. For Sparse, SF and DF one can obtain a further polynomial asymptotic improvement in qubit and Toffoli counts by using Wannier or Bloch functions if the system possesses translational symmetry (see Ref.\citenum{ivanov_quantum_2023} for Sparse algorithm with Wannier and Bloch functions, Ref.\citenum{rubin_fault-tolerant_2023} for Sparse algorithm and other factorizations with Bloch functions). For the Sparse approach, $S=O(N_A^4)$ and for Double Factorization, $\Xi=O(N_A)$ in the worst case scenario. Analytical and numerical results suggest that, $\lambda_{2} = O(N_b^x) = O(N_A^x)$, with  $ 2 \leq x < 2.5$.}
    \label{tab:scaling_comparison}
\end{table}
\subsection{Quantum Resource Estimates}
\label{sec:quantum resource estimates}
In order to estimate the energy within a given error, one has to account for the different approximations used in both the quantum algorithm and Hamiltonian approximations.  The total error in the single point calculations (one estimation of the energy for a given position of nuclei) can be decomposed in a sum of the following errors:

\begin{equation}
    \epsilon_{\rm tot} = \epsilon_{\rm QPE} + \epsilon_{\rm trunc} + \epsilon_{\rm BE} +   \epsilon_{\rm orb} +  \epsilon_{\rm pw} + \epsilon_{\rm paw}, 
\end{equation}
where $\epsilon_{\rm QPE}$ is error from the QPE measurement, $\epsilon_{\rm trunc}$ is the error due to the truncation of the Hamiltonian matrix elements, $\epsilon_{\rm BE}$ is the error incurred when constructing the block encoding, $\epsilon_{\rm orb}$ is the error due to the finite number of orbitals (the number of natural orbitals, $N_b$), $\epsilon_{\rm pw}$ is the error due to the finite size of the plane wave basis,  and $\epsilon_{\rm paw}$ is the error due to the PAW approximation. $\epsilon_{\rm BE}$ is affected by the bitlengths of various parameters used in the circuit: The parameters $\aleph$ for the bitlength of keep probabilities affect the error in amplitudes from coherent alias sampling, and the parameters $\beth$ affects the error in the rotations~\cite{lee_even_2021}. Usually $\epsilon_{\rm BE}$ is negligible compared to the other errors; consistent with the OpenFermion~\cite{mcclean2020openfermion} implementation of double factorisation costing as in~\cite{lee_even_2021}, we simply use the values $\aleph=10$ and $\beth=20$ throughout. Below, we will analyze the other errors and provide our resource estimates for crystalline solids and the defect states.

\subsubsection{Crystalline solids}
In this section, we present the quantum resource requirements for estimation of the ground state energy of crystalline solids without defects. Our goal is to estimate the energy per cell consisting of many atoms and using large basis sets. In order to estimate such an energy at the large basis set limit, we use the down-sampling method. While this method is usually applied w.r.t to $k$-point sampling, it also can be applied with respect to supercell size and, in fact, when the $k$-point mesh is $\Gamma$-centred both approaches are equivalent. To estimate the energy per cell using a $[n+1,n+1,n+1]$ supercell, one can estimate the converged energy per cell for $[n, n, n]$ and add the energy difference between $[n+1, n+1, n+1]$ and $[n, n, n]$ for a fixed number of bands. Let 
$E(n, n_b)$ denote the ground state energy estimated using  a $[n, n, n]$ supercell with $n_b=N_b/N_A$ bands per atom. Then the down-sampling energy is
\begin{equation}
    E_{\rm ds}(n+1) = E_{\rm ds}(n) + E(n+1, n_b) - E(n, n_b),
\end{equation}
We apply this equation up to $n=3$ which results in:
%
\begin{align}
    E_{\rm ds}(1) &= E(1, n_b'')  \\ 
    E_{\rm ds}(2) &= E_{\rm ds}(1) + E(2, n_b') - E(1, n_b') = E(1, n_b'') + E(2, n_b') - E(1, n_b')\\
    E_{\rm ds}(3) &= E_{\rm ds}(2) + E(3, n_b) - E(2, n_b) \nonumber \\ 
    &= E(1, n_b'') + E(2, n_b') - E(1, n_b') + E(3, n_b) - E(2, n_b),
\end{align}
%
where the three parameters can be chosen $n_b< n_b'< n_b''$ to control the number of orbitals per atom and the error in the energy, $\varepsilon_{\rm orb}$. Therefore, one would have to run 5 QPE calculations and the total error for $E_{\rm ds}(3)$ would be a sum of 5 errors from each single point calculations.

To reduce the amount of classical data from QROAM, it is a common practice to truncate the small values of the parameters which determine the Hamiltonian elements~\cite{von_burg_quantum_2021,lee_even_2021} and this affects $\epsilon_{\rm trunc}$. The most dominant cost is due to the plane wave expansion of the Coulomb kernel and the factorization of the orbital-pair density matrices. Therefore, we truncate the inner rank of the plane-wave decomposed two-body term, namely, we set $f_{p,j}(\tG)$ to zero if
\begin{equation}
   | f_{p,j}(\tG) | \leq \delta |\tG|,
\end{equation}
where $\delta$ is a small number. Therefore, $\epsilon_{\rm trunc}$ depends on $\delta$, and so $\epsilon_{\rm trunc} = \epsilon_{\rm trunc}(\delta)$.
Below, we will identify all error parameters and we start with the quality of the UPAW setups and plane wave basis, $\epsilon_{\rm paw}$ and $\epsilon_{\rm pw}$.

In order to estimate the last two errors, $\epsilon_{\rm pw}$ and $\epsilon_{\rm paw}$, we will use density functional theory where we can compare the result of our calculations to both the high-quality all-electron calculations for solids ($\Delta$-DFT data set~)~\cite{delta_dft, lejaeghere_error_2014, lejaeghere_reproducibility_2016} as well as simple diatomic molecules. To generate all-electron binding curves for molecules, we used the uncontracted ANO-RCC-VQZP Gaussian basis set~\cite{roos_main_2004}, readily available from Basis Set Exchange~\cite{pritchard_new_2019,schuchardt_basis_2007,feller_role_1996}, and scalar relativistic corrections as implemented in PySCF~\cite{sun_recent_2020,sun_pyscf_2018,sun_libcint_2015}.  We consider PAW setups which are available in GPAW~\cite{mortensen_gpaw_2023,enkovaara_electronic_2010,mortensen_real-space_2005}, as well as new UPAW setups, norm-conserving Vanderbilt pseudopotentials~\cite{sg15_web,hamann_optimized_2013} and HGH pseudopotentials (a relativistic version of GTH)~\cite{goedecker_separable_1996,hartwigsen_relativistic_1998,krack_pseudopotentials_2005}. Results on norm-conserving pseudopotentials for $\Delta$-DFT data are readily available from Ref.\citenum{delta_dft} and~\citenum{sssp,prandini2018precision}.
All plane-wave basis set calculations have been carried out with GPAW~\cite{mortensen_gpaw_2023,enkovaara_electronic_2010,mortensen_real-space_2005} and when we simulated solids we have used a dense  Monkhorst-Pack k-point grid~\cite{monkhorst_special_1976} (16 points/\AA$^{-1}$).

\begin{figure}  
    \centering
    \includegraphics[width=0.9\linewidth]{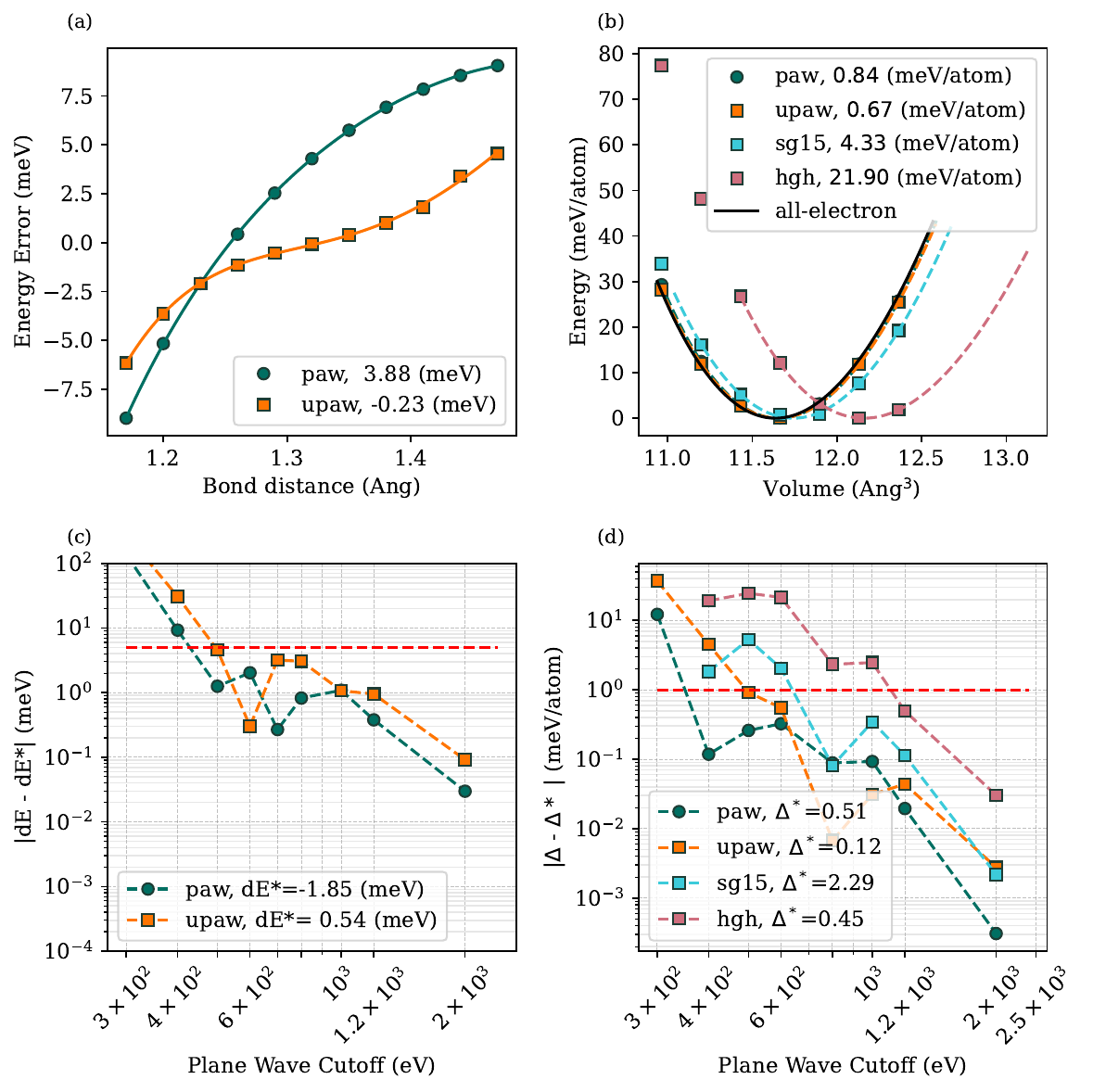}
    \caption{ 
    {\bf Accuracy of norm-conserving pseudopotential and PAW approaches for Carbon.}  (a) C$_2$ energy error curve calculated as the difference between binding curves of all-electron approach using Gaussian type orbitals and (U)PAW setups with plane waves. The kinetic energy cutoff is 600 eV. Labels also indicate the binding energy error (energy difference at the minima) w.r.t. to all-electron calculations. (b) Equation of state of graphite calculated with norm-conserving pseudopotentials (SG15~\cite{sg15_web,hamann_optimized_2013}, HGH~\cite{goedecker_separable_1996,hartwigsen_relativistic_1998,krack_pseudopotentials_2005}), PAW and UPAW setups at 600 eV plane-wave cutoff. All-electron data is from $\Delta$-DFT data set~\cite{delta_dft, lejaeghere_error_2014, lejaeghere_reproducibility_2016}. Labels also indicate the error, $\Delta$, w.r.t. to all-electron calculations. $\Delta$ is defined as the root-mean-square energy difference between the equations of states obtained with all-electron calculations and pseudopotentials/(U)PAW approaches.
    (c) Convergence of binding energy error towards the high-plane wave cutoff limit of 2500 eV. The red dotted line corresponds to an error of 5 meV (0.18 mHa). (d) Convergence of error, $\Delta$, towards the high-plane wave cutoff of 2500 eV. The red dotted line corresponds to an error bar of 1.0 meV/atom. The lower the curve the closer the result to the converged values.
    }
    \label{fig:C_upaw_accuracy}
\end{figure}

Fig.~\ref{fig:C_upaw_accuracy} shows the results of these calculations. Fig.~\ref{fig:C_upaw_accuracy}(a) shows the binding curves obtained at 600 eV with UPAW and PAW setups. As can be seen the error in the binding energy is well within chemical accuracy. 600 eV corresponds to practically converged results as shown in Figure~\ref{fig:C_upaw_accuracy}(c). The PAW method demonstrates slightly faster convergence (around 400 eV) as expected. Fig.~\ref{fig:C_upaw_accuracy}(b) shows the equation of states of graphite calculated with different methods at the small value of 600 eV, and Fig.~\ref{fig:C_upaw_accuracy}(d) demonstrates the convergence of error w.r.t all-electron calculations towards the limit of 2500 eV. The lower the curve on that graph the closer the result to the converged value. The converged values with UPAW and PAW can be achieved already at 600 eV while norm-conserving pseudopotentials require a higher plane-wave cutoff. While HGH pseudopotential demonstrates good accuracy for this material, it requires the largest plane wave cutoff among all methods. For example, to reach an accuracy of 1.0 meV/atom one has to use more than 1000 eV kinetic-energy cutoff and for an accuracy of 0.1 meV/atom, one has to use even higher a 1500-2000 eV cutoff. We also note that we do not guarantee that 2500 eV corresponds to absolute convergence, but higher plane wave cutoffs are too computationally expensive to use in any practical calculations. As can be seen, the error introduced due to the (U)PAW approximation and finite PW basis set is small, -0.12 meV/atom $\approx$ 0.044 mHa/atom, and in the following resource estimations we will use a kinetic-energy cutoff of 600 eV. In order to show that the UPAW approach is transferable to other more challenging elements such as transition metals, we present the results for a few other elements based on the $\Delta$-DFT data in the Table~\ref{tab:comparison_ps}. As one can see, the norm-conserving pseudopotentials exhibit large errors for elements such as Cr and Mn, while the PAW-based methodology provides a small error. A particularly difficult system is Cr bulk where the inclusion of semicore electrons (3$s$ and 3$p$) is necessary in UPAW setups in order to obtain accurate results. However, UPAW setups with a small number of electrons (6 valence electrons) for Cr is comparable in accuracy with norm-conserving pseudopotentials where semicore electrons are included. We note, however, that it is probably possible to refit the norm-conserving pseudopotentials to minimize this error.

\begin{figure} 
    \centering
    \includegraphics[width=0.9\linewidth]{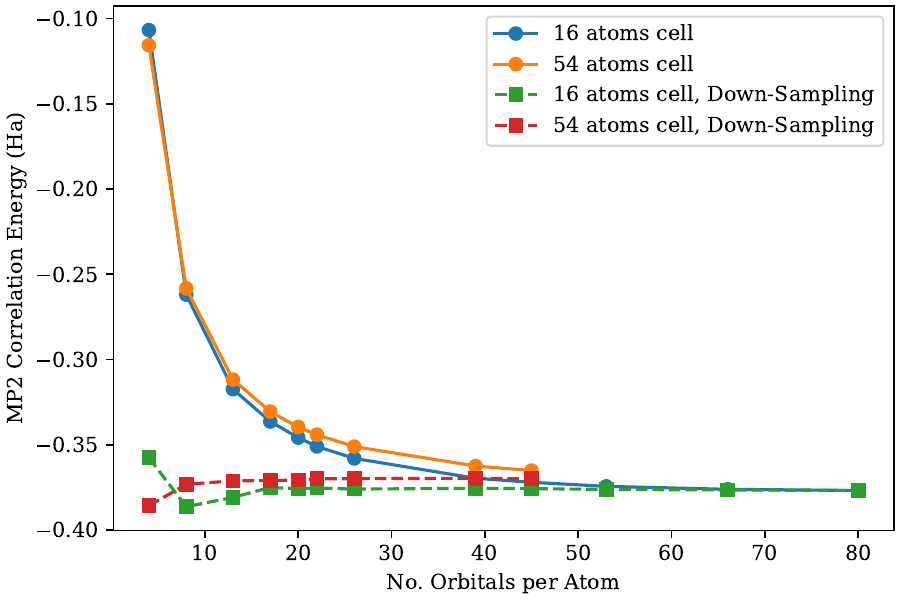}
    \caption{{\bf Convergence of MP2 correlation energy per cell (2 atoms) of diamond with respect to the number of orbitals per atom.} Calculations are presented for both the down-sampling method and the supercell approach (16 and 54 atoms cells).}
    \label{fig:mp2_correlation_energy}
\end{figure}

To choose the number of orbitals per atom, $n_b, n_b', n_b''$, as well as the truncation threshold $\delta$, the MP2 correlation energy is calculated for different numbers of orbitals until convergence is reached up to the desired accuracy. Fig.~\ref{fig:mp2_correlation_energy} and Table~\ref{tab:MP2 correlation error} show results of such calculations using both a standard supercell approach and down-sampling method. As is expected, the down-sampling approach converges much faster than the supercell approach and reaches chemical accuracy already at 13 orbitals/atom. We then carry out calculations with $n_b''$ and different values of $n_b', n_b$ and $\delta$ to find allowable error parameters within the error budget. We find that with $\delta=3\cdot10^{-5}$ and $n_b'=26$, we stay within chemical accuracy for $n_b\geq 17$, and the error is $0.96$ mHa. Consequently, we have to estimate the quantum resources for running QPE with 5 sets of parameters as presented in Table~\ref{tab:resource estimates}. For each set, we estimate the quantum resources with QPE accuracy of the remaining error budget, $\epsilon_{\rm QPE} = (1.60 - 0.96 - 0.44)/5 \approx 0.04 $ mHa. The value of $0.44$ corresponds to estimation from UPAW and finite plane-wave basis set as was described above. As can be seen the most computationally demanding calculations are for the largest system with a 54 atom cell and 17 orbitals per atom. This requires around 140,000 logical qubits and a number of Toffolis on the order of $10^{14}$. We note that even smaller calculations with 2 atoms in the cell and 160 orbitals/atoms require large quantum resources, $\approx 13,000$ logical qubits and on the order of $10^{13}$ Toffolis. Assuming that this computation will be run on a superconducting device with nearest-neighbour connectivity on a square lattice with a physical error rate of $0.01\%$, we estimate that such calculations would require around 23.8 million and 316 million physical qubits for 2 atoms cell with 80 orbitals per atom and 54 atom cell with 17 orbitals per atom, respectively. This estimation assumes surface code error correction (see Appendix~\ref{app:error correction} for more details).

\begin{table}
    \centering
    \begin{tabularx}{1.0\textwidth} { 
   >{\centering\arraybackslash}X 
   >{\centering\arraybackslash}X 
   >{\centering\arraybackslash}X    
   >{\centering\arraybackslash}X 
   >{\centering\arraybackslash}X 
  }
    \hline
        Element &  UPAW & PAW & SG15~\cite{sssp}& HGH~\cite{delta_dft} \\
    \hline
         O & 0.55 & 0.17 & 0.39 & 1.30 \\
         Cr & 0.47 (12.6) & 3.10 & 20.82 & 13.68 \\
         Mn & 0.67 (2.85) & 1.01 & 13.05 & 15.68\\
         Ni & 0.54 (0.18) & 4.51 & 2.16 & 1.38 \\
    \end{tabularx}
    \caption{ {\bf The error, $\Delta$, w.r.t. to all-electron calculations~\cite{delta_dft, lejaeghere_error_2014, lejaeghere_reproducibility_2016}.} $\Delta$ is defined as the root-mean-square energy difference between the equations of states obtained with all-electron calculations and a tested approach. Values are in the units of meV/atom. Values in parentheses are obtained from setups with only valence electrons (6, 7, 10 electrons for Cr, Mn and Ni, respectively). All other calculations include semicore electrons explicitly.}
    \label{tab:comparison_ps}
\end{table}

\begin{table}
    \centering
    \begin{tabularx}{1.0\textwidth} { 
   >{\centering\arraybackslash}X 
   >{\centering\arraybackslash}X    
   >{\centering\arraybackslash}X 
  }
    \hline
    Orbitals per Atom & $\epsilon_{\rm orb}$ & $\epsilon_{\rm orb}  + \epsilon_{\rm trunc}(\delta=3\cdot 10^{-5})$ \\
    $n_b$ & ($n_b'', n_b'$) = (80, 80)& ($n_b'', n_b'$) = (80, 26)\\
    \hline
    4 &  15.80 &  16.93\\
    8 &  3.25 &  3.87\\
    13 &  1.25 &  2.40\\
    17 &  1.08 &  0.96\\
    20 &  0.72 &  0.17\\
    22 &  0.02 & -0.60\\
    26 & -0.04 & -0.70\\
    \hline
    \end{tabularx}
    \caption{{\bf Error due to the finite size of the orbital basis set and truncation}. Deviation of the MP2 correlation energy from an accurate value estimated using down-sampling technique with $n_b=45$ orbitals per atom and a (3, 3, 3) supercell with $n_b'' = n_b'=80$. The last column presents the deviation for a finite truncation and for $n_b''=80$ and $n_b'=26$. Energy is given in mHa/primitive cell (2 atoms).}
    \label{tab:MP2 correlation error}
\end{table}
\begin{table}
    \centering
    \begin{tabularx}{1.0\textwidth} { 
   >{\centering\arraybackslash}X 
   >{\centering\arraybackslash}X 
   >{\centering\arraybackslash}X 
   >{\centering\arraybackslash}X 
  }
    \hline
    Size of the cell & Orbitals per Atom & Logical Qubits & Toffolis\\
    $(n, n, n)$ & &  & \\
    \hline
    (1, 1, 1) & $n_b'' = 80$ & 13313 & $7.33 \cdot 10^{13}$\\
    (1, 1, 1) & $n_b' = 26$ & 4443 & $2.51 \cdot 10^{12}$\\
    (2, 2, 2) & $n_b' = 26$ & 67593 & $1.85 \cdot 10^{14}$ \\
    (2, 2, 2) & $n_b = 17$ & 44262 & $4.41 \cdot 10^{13}$\\
    (3, 3, 3) & $n_b = 17$ & 148937 & $5.23 \cdot 10^{14}$\\
    \hline
    \end{tabularx}
    \caption{\textbf{Quantum resource estimates for diamond.} The QPE error budget is 0.04 mHa/primitive cell.}
    \label{tab:resource estimates}
\end{table}

\subsubsection{Defects in Solids: Nitrogen-Vacancy Centre in Diamond.} 
Our methodology can be applied to materials with defects. Typical calculations of defects are performed in the supercell approach without $k$-point sampling, and the size of the supercell should be large enough so as to reduce the interaction between periodic images of the defect state. Here we estimate the resources for energy state estimation using the supercell approach.

\begin{figure}
    \centering
    \includegraphics[width=1.0\linewidth]{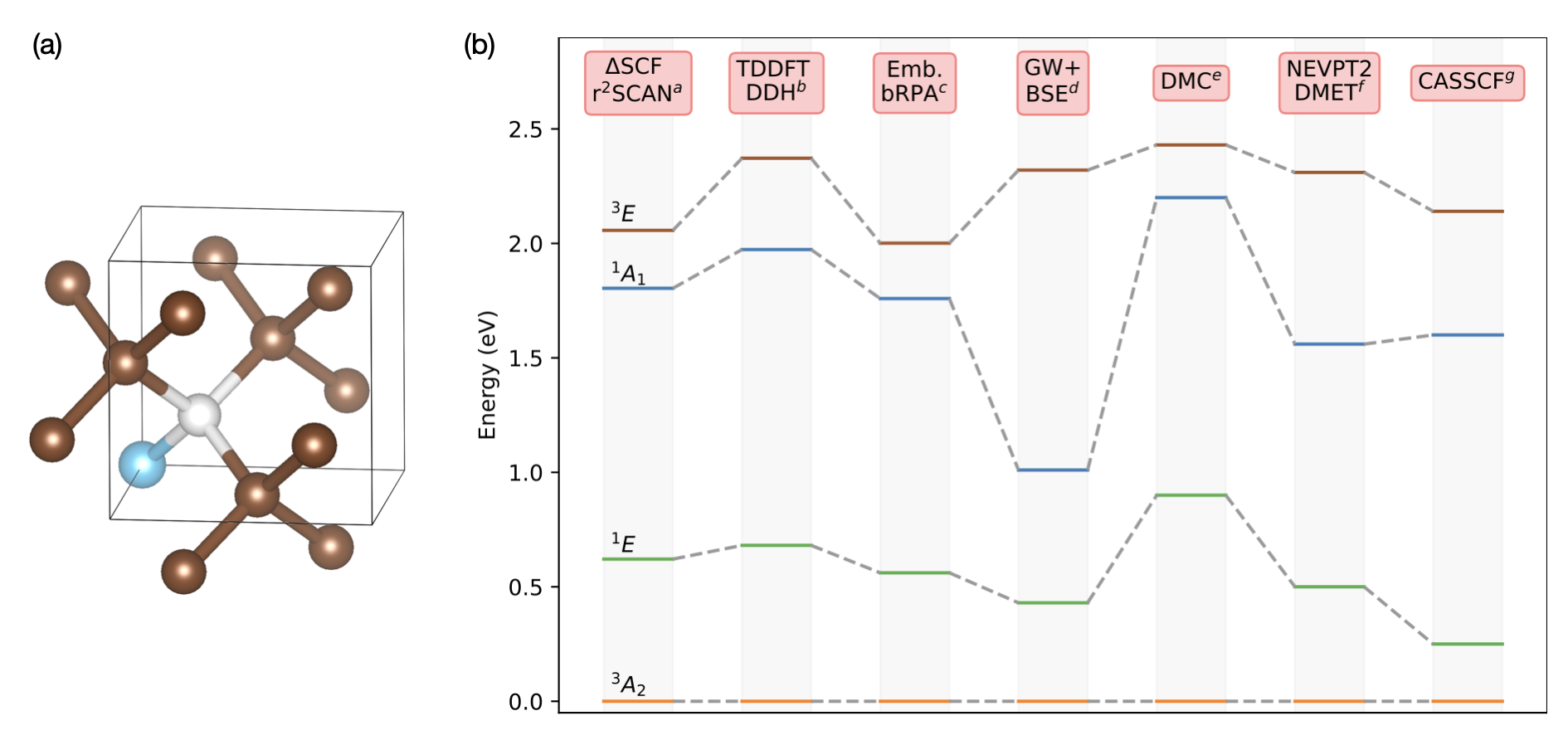}
    \caption{\textbf{Nitrogen-vacancy centre in diamond and excitation energies} (a) Schematic representation of the centre. Carbon atoms are brown, Nitrogen is blue, and the vacancy is white (b) Excitation energy of the negatively-charged NV centre calculated using different classical algorithms. ${}^{a}$Ref.\citenum{ivanov_electronic_2023}, ${}^{b}$Ref.\citenum{Jin2022}, ${}^{c}$Ref.\citenum{ma_quantum_2020}, ${}^{d}$Ref.\citenum{ma_excited_2010}, ${}^{e}$Ref.\citenum{simula_calculation_2023}, ${}^{f}$Ref.\citenum{haldar_local_2023}, ${}^{g}$Ref.\citenum{bhandari_multiconfigurational_2021}.
    Figure (b) is modified version of the figure 3 from Ref.\citenum{ivanov_electronic_2023} (available under the terms of the Creative Commons Attribution 4.0  license Copyright 2023)
    }
    \label{fig:nv-centre}
\end{figure}

Quantum defects in semiconductors are of great interest as they can be utilized in a range of applications such as sensing~\cite{schirhagl_nitrogen-vacancy_2014, maze_nanoscale_2008, balasubramanian_nanoscale_2008, taylor_high-sensitivity_2008}, quantum communication~\cite{hensen_loophole-free_2015, aharonovich_diamond-based_2011, beveratos_single_2002} and computation~\cite{waldherr_quantum_2014, taminiau_universal_2014, doherty_nitrogen-vacancy_2013, weber_quantum_2010, neumann_multipartite_2008}. A negatively charged Nitrogen-Vacancy (NV$^{-}$) centre in diamond is one of the most studied and understood defects. However, predicting the excitation energy levels using electronic structure methods is challenging due to large supercells required to reduce the interaction of the centre with its periodic images (the length-scale problem) as well as the multi-determinantal nature of singlet excited states. 
Fig.~\ref{fig:nv-centre} shows energy levels of the excited triplet ${}^{3}E$ and two singlet states,
${}^{1}E$ and ${}^{1}A_1$,  obtained with different electronic structure methodologies~\cite{ivanov_electronic_2023,Jin2022,ma_quantum_2020,ma_excited_2010,simula_calculation_2023,haldar_local_2023,bhandari_multiconfigurational_2021}. The most challenging state is the singlet ${}^{1}A_1$ state and as one can see the prediction of its energy ranges from $\sim\!1.1$ eV (GW-BSE calculations~\cite{ma_excited_2010}) up to $\sim\!2.1$ eV (Diffusion Monte Carlo calculations~\cite{simula_calculation_2023}). 
Given such a large scattering of the data as given by mean-field approaches, many-body perturbation theory, quantum embedding and wavefunction methods, it is clear that having an additional more reliable method to predict the energetics of such a system would be very beneficial. We note that such and similar systems have been explored using quantum embedding methods in the context of quantum computation~\cite{ma_quantum_2020,baker_simulating_2024}. However, the quantum embedding methods used there rely on many approximations such as an exchange-correlation functional, the way in which the double-counting term is implemented, and the level of theory used for calculation of the screened Coulomb interaction~\cite{muechler_quantum_2022}. These approximations introduce uncontrollable errors in the calculations and results can then be verified only after the comparison with experimental data is made. Calculations in the supercell approach without additional approximations would be more predictive.

\begin{figure}
    \centering
    \includegraphics[width=1.0\linewidth]{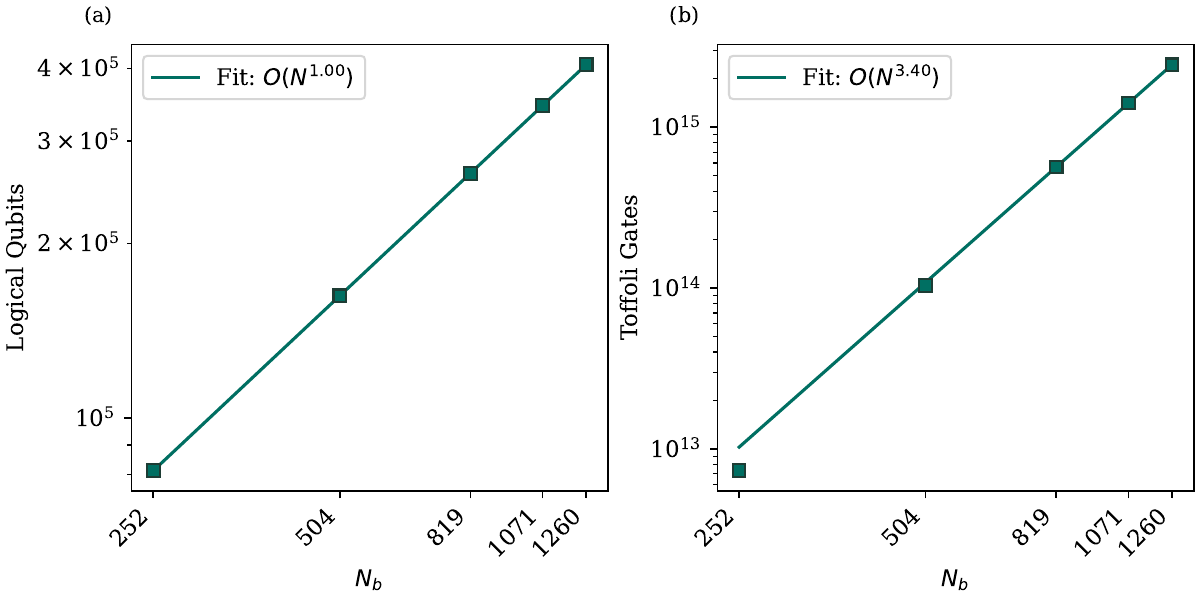}
    \caption{{\bf The number of logical qubits (a) and Toffoli gates (b) vs. the number of orbitals required for QPE computation of the ground state of NV$^{-}$ centre in diamond.} Calculations have been carried out for a 63 atom supercell and the QPE error budget is 1.0 mHa.}
    \label{fig:nv-centre-qandt} 
\end{figure}

Fig.~\ref{fig:nv-centre-qandt} shows the quantum resource requirements for the energy estimation of this system on an error-corrected quantum computer.
In the calculations, we used 4, 8, 13, 17 and 20 natural orbitals per atom.
Since we do not carry out the error analysis due to truncation using MP2 theory (because of the degenerate homo-lumo gap), we choose the conservative truncation parameter $\delta = 3 \cdot 10^{-6}$, an order of magnitude lower than in the previous section. The QPE error budget is 1.0 mHa per supercell. We have used a 600 eV kinetic-energy cutoff and UPAW setups as described before.
Fig.~\ref{fig:nv-centre-qandt} shows that even the smallest instances of such simulations would require around 80,000 logical qubits and the number of Toffolis is around $7.3\cdot 10^{12}$. With the increase of the number of orbitals, the number of logical qubits grows only linearly and the Toffoli count, which would be proportional to the total runtime, scales polynomially as $O(N_b^{3.4})$.

\section{Discussion and Conclusion}\label{sec:discussion_and_conclusion}

We have introduced the unitary projector augmented-wave method and used it with a plane-wave basis set to derive an efficient representation of Hamiltonians suitable for quantum computation of materials. We have analysed the resource requirements of this approach for high-accuracy calculations.  We note that our implementation is also valid for norm-conserving pseudopotentials since the latter can be implemented as a special case of the PAW method~\cite{mortensen_gpaw_2023}. While there have been recent developments in quantum computing methods for materials calculations using norm-conserving pseudopotentials and Gaussian basis sets~\cite{ivanov_quantum_2023, rubin_fault-tolerant_2023}, it is somewhat hard to compare the approaches because the comparison of quantum resources should be made for the same accuracy of calculations. To achive this, one would need to carry out all-electron correlated calculations of materials at scale and converged basis set limits and produce reliable reference values. Such benchmarks are available for density functional theory~\cite{lejaeghere_reproducibility_2016,prandini2018precision,bosoni_how_2023} but we are not aware of any for wavefunction calculations. Such simulations would require enormous CPU time and we did not attempt to carry out this comparison in this work. However, if we take a number of natural orbitals in our calculations that corresponds to the size of the cc-pVDZ basis set (13 orbitals per atom) for diamond and estimate the quantum resources with a QPE error budget of 1.6 mHa/cell, we obtain $4.0 \cdot 10^{11}$ Toffolis and 33878 logical qubits, which is similar to the value obtained using tensor-hypercontraction in Ref.\citenum{rubin_fault-tolerant_2023} ($4.85 \cdot 10^{11}$ Toffolis and 36393 logical qubits). We note that using Bloch functions with $k$-point sampling and primitive cell instead of supercell calculations~\cite{ivanov_quantum_2023, rubin_fault-tolerant_2023} can provide a polynomial speed up for crystalline solids. However, this does not bring any advantage for other applications such as defects states where calculations must be carried out in the supercell so as to minimize the interaction between periodic images. For this reason, we did not explore Bloch or Wannier functions in this work. Similarly to classical methodologies, our approach is expected to be particularly beneficial when the plane wave basis set is more efficient than Gaussian orbitals. This includes, but is not limited to, applications such as high-precision calculations of dense systems and the calculation of optical properties. For the latter, the unbiased description of ground and excited states offered by plane waves proves particularly advantageous for, e.g., determining band structures or accurately describing a broad excitation spectrum, avoiding the need for the highly specialised Gaussian basis sets. Furthermore, the accurate calculation of energy derivatives in plane wave basis set (e.g., forces without basis set superposition error, which is a principal error source in atomic orbital basis sets such as Gaussians), mean that exploring UPAW with plane waves in the context of quantum algorithms for energy derivatives is a promising research area.

We have also presented a strongly correlated system, the negatively charged nitrogen-vacancy defect in diamond, where classical algorithms provide scattered results for the the energy of the excited states.
Investigating such systems is especially valuable because, in the challenging search for problems with the quantum advantage~\cite{lee2023evaluating}, strongly correlated systems~\cite{izsak2023measuring} remain the best candidate. While the quantum computing algorithms require resources that scale polynomially with system size, significant advances must be made to reduce the quantum resources further to a feasible size for realistic fault-tolerant quantum computers. Only then could one potentially use such algorithms for accurate \textit{ab initio} calculations of the defect states. There are promising avenues for further reducing resource costs, by considering the electronic structure problem in first quantization instead of second quantization. Interesting works on incorporating  the Goedecker-Teter-Hutter norm-conserving pseudopotentials~\cite{goedecker_separable_1996,hartwigsen_relativistic_1998} in first quantization have recently been published~\cite{zini_quantum_2023, berry_quantum_2023}. The PAW method quite often requires a smaller number of plane waves and demonstrates a higher accuracy as indicated in density functional theory calculations~\cite{delta_dft,lejaeghere_reproducibility_2016,prandini2018precision,bosoni_how_2023}, especially for transition metals as is shown in Table~\ref{tab:comparison_ps}. Therefore, it will be beneficial to also implement the PAW approach within first quantization. Unlike pseudopotentials, the PAW approach also modifies the two-body term and in order to incorporate this, non-trivial modifications to the method presented in Refs.~\citenum{berry_quantum_2023,su_yuan_fault-tolerant_2021,babbush_sublinear_2019} might be required. One simple way to incorporate the PAW method is to use the first quantization approach developed in Ref.~\citenum{georges_quantum_2024}, which loads the electronic integrals from QROAM, unlike the approach of Refs.~\citenum{su_yuan_fault-tolerant_2021,babbush_sublinear_2019}. By additionally employing basis sets that diagonalize the two-body term, such as dual plane waves~\cite{babbush_low-depth_2018}, such an approach might provide a viable path to significant reduction of quantum resources. 

\begin{acknowledgement}
We thank Nick Blunt for discussions, carefully reading the manuscript and providing valuable suggestions.
The work presented in this paper was part funded by a grant from Innovate UK under the `Feasibility Studies in Quantum Computing Applications' competition (Project Number 10074148).
M.B. is a Sustaining Innovation Postdoctoral Research Associate at Astex Pharmaceuticals and thanks Astex Pharmaceuticals for funding, as well as his Astex colleague Patrick Schoepf for his support.
\end{acknowledgement}

\appendix

\section{Matrix elements in PAW formalism and plane wave basis set}\label{sec: matrix elements}
Let $\gamma^{a}$ be the number of core electrons, $\zeta^{a}_j(\tr)$ the $j\textsuperscript{th}$ core orbital, $\nu^{a}(\tr)$ the core electron density, and $\mathcal{Z}_{a} = Z_a \delta(\tr - \tP_a)$ the nuclear charge density of the atom $a$. The constant term in the Hamiltonian is 
\begin{equation}
         H^{(0)} =  2 \sum_{a=1}^{N_A}\sum_{j=1}^{\gamma^{a}/2} \braket{\zeta^{a}_j|-\frac{1}{2}\nabla^2 |\zeta^{a}_j}  
        +
        \sum_{aa'}{\vphantom{\sum}}'\frac{1}{2}(\nu^{a} + \mathcal{Z}_{a}|\nu^{a'} + \mathcal{Z}_{a'})  - \sum_{aa'}\sum_{jk}^{\gamma^{a}/2} (\zeta^{a'}_k \zeta^{*a}_j| \zeta^{*a'}_k \zeta^{a}_j),
\end{equation}
where prime over the sum indicates that the self-interaction energy is not included, and where we introduced the notation
\begin{equation}
    (f|g) = \IINTr{} \frac{f^*(\tr') g(\tr)}{|\tr' - \tr|}.
\end{equation}
The one-body matrix elements consist of the valence electron kinetic and external potential contributions, the interaction of valence electrons with core electrons, and the PAW correction:
\begin{multline}    
h_{p q} = - \frac{1}{2}
\INTr{V} \tilde{\psi}^{*}_{p}(\tr) \nabla^2 \tilde{\psi}_{q}(\tr) - 
\sum_{a=1}^{N_A} Z_{a}
\INTr{V} \frac{\tilde{\rho}_{pq}(\tr)}{|\tr - {\bf P}_a|} + \\
\sum_{a=1}^{N_A}
\frac{\gamma^{a}}{\sqrt{4\pi}}\left(\tilde{g}^{a}_{0}\right | \tilde{\rho}_{pq}) +
\sum_{a=1}^{N_A}
\sum_{i_1 i_2}^{n_a} D^{a}_{pq, i_1 i_2} 
\left[H^{a}_{i_1i_2} + V^{a}_{i_1i_2} - X_{i_1i_2}^{a},
\right]
\end{multline}    
where
\begin{equation}
    \tilde{\rho}_{pq}(\tr) = \tilde{n}_{pq}(\tr) + \sum_{a=1}^{N_A} \tilde{Z}_{pq}^{a}(\tr), \quad \tilde{n}_{pq}(\tr) = \tilde{\psi}^{*}_{p}(\tr) \tilde{\psi}_{q}(\tr),
\end{equation}
\begin{equation}
    \tilde{Z}^{a}_{pq}(\tr)  = \sum_{L=0}^{L_{\rm max}} Q^{a}_{pq, L} \tilde{g}^{a}_L(\tr), \quad      
    Q^{b}_{pq, L} = \sum_{i_1 i_2}^{n_a} \Delta_{Li_1 i_2}^{a} D_{pq,i_1i_2}^{a}.
\end{equation}
where $D_{pq,i_1i_2}^{a}$ is the atomic orbital-pair density matrix:
\begin{equation}
    D_{pq,i_1i_2}^{a} = \braket{\tilde{\psi}_p|p^{a}_{i_1}} \braket{p^{a}_{i_2}|\tilde{\psi}_q}
\end{equation}
The atomic compensation charges $\sum_{a} Z_{pq}^{a}(\tr)$ are introduced to ensure that the Coulomb potential created by atomic-centered densities are zero outside the augmentation spheres, which allows for separation of the original Hamiltonian into soft and atomic parts only. The same approach will be used below for the two-body term. The atomic constants due to the kinetic energy and external potential contribution, $H_{i_1i_2}^{a}$, the Hartree energy of valence and core electrons, $V_{i_1i_2}^{a}$, and the exchange energy between valence and core electrons, $X_{i_1i_2}^{a}$, are
\begin{equation}
    H^{a}_{i_1i_2} = \braket{\phi_{i_1}^a| -\frac{1}{2} \nabla^2 - \frac{Z_a}{|\tr - \tP_a|} |\phi_{i_2}^a} - \braket{\tilde{\phi}_{i_1}^a| -\frac{1}{2} \nabla^2 - \frac{Z_a}{|\tr - \tP_a|} |\tilde{\phi}_{i_2}^a}.
\end{equation}
\begin{equation}
V_{i_1i_2}^a = 
 \left(\phi^{a}_{i_1} \phi^{a}_{i_2} | n^{a,\core}\right) - 
 \frac{\gamma^{a}}{\sqrt{4\pi}} \left(\tilde{\phi}^{a}_{i_1} \tilde{\phi}^{a}_{i_2}| \tilde{g}^{a}_{0}\right)
 -
 \frac{\gamma^{a}}{\sqrt{4\pi}} (\tilde{g}_{0}^{a}|\tilde{g}_{0}^{a}) \Delta_{0,i_1 i_2}^{a}, 
\end{equation}
\begin{equation}
    X_{i_1i_2}^a = \sum_{j=1}^{\gamma^{a}/2} (\phi_{i_1}^{a} \zeta_{j}^{a}|\phi_{i_2}^{a}\zeta_{j}^{a}),
\end{equation}
and $\tilde{g}_L^{a}(\tr)$ is a Gaussian function localized on atom $a$, with angular and magnetic numbers $L=(l, m)$. We note that there are approaches which parameterize the all-electron one-body term~\cite{schmerwitz_revisiting_2024} in different way using Fock matrix, but the approach of Ref.\citenum{schmerwitz_revisiting_2024} requires implementation of the additional term that accounts for double-counting. 
The two-body term is
\begin{equation}
    \ka_{pqrs} = (\tilde{\rho}_{pq}|\tilde{\rho}_{rs}) +  \sum_{a}^{N_A}
    \sum_{i_1i_2i_3i_4}^{n_a} C^{a}_{i_1 i_2 i_3 i_4} D_{pq,i_1i_2}^{a*} D_{rs,i_3i_4}^{a} \, .
\end{equation}
The matrix element is expanded into a sum of the soft contribution and atomic PAW correction. The definition of the coefficient $C^{a}_{i_1 i_2 i_3 i_4}$ is given in Ref.\citenum{rostgaard_projector_2009}. The most computationally intensive part for calculating matrix elements for such a Hamiltonian is the soft two-body contribution since it is a 4-rank tensor. We will use a plane-wave basis set in order to derive an LCU decomposition of the two-body term.

\subsection{Factorization of soft two-body term using plane waves}
The soft part of the two-body term can be expanded as:
\begin{equation}    
    \tilde{\kappa}_{pqrs} = (\tilde{\rho}_{pq}|\tilde{\rho}_{rs}) = \sum_{j} (\eta_{pq,j}|\eta_{rs,j})
\end{equation}
where 
\begin{equation}
    \eta_{pq,0}(\tr) = \frac{\tilde{\rho}_{pq}(\tr) + \tilde{\rho}_{pq}(-\tr)}{2}, \quad  \eta_{pq,0} (\tr) = \eta_{pq,0}(-\tr) = \eta_{qp,0}^{*}(\tr)
\end{equation}
and
\begin{equation}
    \eta_{pq,1}(\tr) = \frac{\tilde{\rho}_{pq}(\tr) - \tilde{\rho}_{pq}(-\tr)}{2i}, \quad \eta_{pq,1}(\tr) = -\eta_{pq,1}(-\tr) = \eta_{qp,1}^{*}(\tr).
\end{equation}
It is convenient to introduce the orbital-pair densities, $\eta_{pq,j}$, because their plane wave coefficients (defined in Appendix~\ref{app: plane-wave expansion}) are Hermitian matrices with reflection and anti-reflection symmetry, respectively:
\begin{align}    
    C_{pq, j}^{*}(\tG) = C_{qp, j}(\tG), \quad C_{pq, j}(-\tG) = (-1)^{j}C_{pq, j}(\tG),
\end{align}
unlike the plane-wave coefficients of the orbital-pair density matrix, $\rho_{pq}(\tr)$, which satisfy the following:
\begin{equation}
    C_{pq}^{*}(\tG) = C_{qp}(-\tG).
\end{equation}
Then, we use the plane-wave expansion of the Coulomb potential
\begin{equation}
    \frac{1}{|\tr|} = \sum_{\tG} e^{i\tG\tr} v({\tG})
\end{equation}
to derive
\begin{equation}
    \tilde{\kappa}_{pqrs} = 
    2
    \sum_{j=1,2} \sum_{\tG \geq 0} v'(\tG) C_{pq,j}^{*}(\tG) C_{rs,j}(\tG),
\end{equation}
with $v'(0) = v(0)/2$, and $v'(\tG)=v(\tG)$ otherwise. As one can see, we do not omit the zero $\tG$ component and instead will use Wigner-Seitz regularization~\cite{sundararaman_regularization_2013}.

The soft two-body term can then be rewritten as follows:
\begin{equation}
    \hat{\tilde{H}}^{(2)} =
    \sum_{j}
    \sum_{\tG\geq0} v'(\tG) \left(\sum_{pq} C_{pq,j}(\tG)\hat{E}_{pq}\right)^2 
\end{equation}
We can further diagonalize $C_{pq,j}(\tG)$ using the fact that for a given $\tG$ it is a Hermitian matrix:
\begin{equation}
    C_{pq,j}(\tG) = \sum_{r} U_{pr,j}(\tG) f_{r,j}(\tG) U^{*}_{qr,j}(\tG),
\end{equation}
then the soft two-body term can be rewritten using free-fermionic unitaries, $\hat U_{j}(\tG)$
as follows~\cite{lee_even_2021}:
\begin{equation}
    \hat{\tilde{H}}^{(2)} = \sum_{\tG \geq 0}^{N_{\rm pw}/2} \sum_{j=1,2}v'(\tG^2) \hat U_{j}(\tG) \left(\sum_{p}^{N_b} f_{p,j}(\tG) \hat E_{pp}\right)^2 \hat U^{\dagger}_{j}(\tG)
\end{equation}

\subsection{Factorization of PAW two-body part}

In order to factorize the PAW two-body term, we use the fact that the atomic Coulomb coefficients $C_{i_1i_2i_3i_4}^{a}$ are symmetric with respect to swapping of a pair of indices $(i_1, i_2)$ and $(i_3, i_4)$. Unlike the electron repulsion integrals, $C_{i_1i_2i_3i_4}^{a}$ are not positive definite and one cannot use Cholesky decomposition which is usually used in the single- and double-factorization methods. Instead one could use a singular-value decomposition but in order to reduce the number of angles that need to be loaded from QROAM, we use regular eigendecomposition and later will simply load the sign of each term in the LCU. 

By introducing the eigendecomposition
\begin{equation}
    \left(\frac{1}{2}\right)^{\delta_{i_1i_2} + \delta_{i_3 i_4}} C^{a}_{i_1i_2i_3i_4} = \sum_{k\leq l} O_{i_1i_2 kl} \epsilon_{kl} O_{i_3i_4 kl}, \quad \text{for}\quad i_1 \leq i_2;\, i_3 \leq i_4,
\end{equation}
we can rewrite the PAW two-body term as follows:
\begin{multline}    
    \hat H^{(2), {\rm PAW}} =
    \sum_{a}^{N_A} \sum_{pqrs}^{N_b}\sum_{i_1i_2i_3i_4}^{n_p^{a
    }} C^{a}_{i_1 i_2 i_3 i_4} D_{pq,i_1i_2}^{a*} D_{rs,i_3i_4}^{a} \hat{E}_{pq}^{\dagger} \hat{E}_{rs} = \\
    \sum_{a}^{N_A}
    \sum_{i_1\leq i_2} \epsilon_{i_1 i_2}^{a}
    \left(\sum_{rs}\sum_{i_3\leq i_4}  O^{a}_{i_3 i_4 i_1 i_2} (D^{a}_{rs,i_3i_4} + D^{a*}_{sr,i_3i_4}) \hat{E}_{rs} \right)^2.
\end{multline}
Since the matrix
\begin{equation}
    L_{pq,i_1i_2}^{a} = \sum_{i_3\leq i_4}  O^{a}_{i_3 i_4 i_1 i_2} (D^{a}_{pq,i_3i_4} + D^{a*}_{qp,i_3i_4}),
\end{equation}
is Hermitian with respect to $p,q$, we can make use of the additional eigendecompostion 
\begin{equation}
    L_{pq, i_1i_2}^{a} = \sum_{r} U^{a}_{pr,i_1i_2} f^{a}_{r, i_1i_2}U^{a}_{qr,i_1i_2}
\end{equation}
and introducing the free fermionic unitaries, $\hat U^{a}_{i_1 i_2}$, we arrive at
\begin{equation}\label{eq:paw_interaction_term}
    \hat H^{(2), {\rm PAW}} = \sum_{a}\sum_{i_1\leq i_2} {\rm sign}(\epsilon_{i_1i_2}^a) \hat U_{i_1i_2}^a\left( \sqrt{|\epsilon_{i_1i_2}^a|} \sum_{r,\sigma} f_{r, i_1i_2}^a \hat n_{r\sigma} \right)^2 \hat U_{i_1i_2}^{a\dagger}
\end{equation}

\section{Properties of reciprocal pair densities}\label{app: plane-wave expansion}
We define the plane wave expansion for a periodic function as
    \begin{equation}
        f(\tr) = \sum_{\tG} e^{i\tG\tr} C(\tG),
    \end{equation}
    where 
    \begin{equation}
        C(\tG) = \frac{1}{V} \INTr{V} e^{-i\tG\tr} f(\tr).
    \end{equation}

\subsection{Complex orbital-pair densities}
For complex-valued orbitals (as well as real), we have the following properties for orbital-pair densities:
\begin{align}
    C^{*}_{pq}(\tG) = C_{qp}(-\tG) \, .
\end{align}
\subsection{Real orbital-pair densities}
Additionally, real orbitals satisfy
\begin{align}
    C_{pq}(\tG) = C_{qp}(\tG) \, ,
\end{align}
which leads to 
\begin{equation}
    C_{pq}(-\tG) = C^{*}_{qp}(\tG) = C^{*}_{pq}(\tG)  \, .
\end{equation}
Thus, we don't need to store the $-G$ component, since the coefficients can be restored by conjugation.
Also, for real orbitals,
\begin{equation}
    \eta_{pq, 0} = \frac{C_{pq}(\tG) + C_{qp}^{*}(\tG)}{2} = \frac{C_{pq}(\tG) + C_{pq}^{*}(\tG)}{2} = {\rm Re}(C_{pq}(\tG)) \, ,
\end{equation}
and 
\begin{equation}
    \eta_{pq, 1} = \frac{(C_{pq}(\tG) - C_{qp}^{*}(\tG))}{2i} = \frac{(C_{pq}(\tG) - C_{pq}^{*}(\tG))}{2i} = {\rm Im}(C_{pq}(\tG)) \, .
\end{equation}

\section{Compression of virtual space with approximate MP2-natural orbitals}\label{sec: mp2no}
In order to reduce the size of the virtual space, we use approximate MP2 natural orbitals
~\cite{aquilante_systematic_2009,gruneis_natural_2011}. These orbitals allow for much faster convergence to the full basis set limit compared to canonical KS or HF orbitals.
By definition, these orbitals diagonalize the approximate-MP2 density matrix~\cite{aquilante_systematic_2009}:
\begin{equation}\label{eq:approx_density_matrix}
    D_{ab} = \sum_{c}^{N_{\rm virt}}\sum_{i}^{N_{\rm occ}}
    \frac{\kappa_{icbi} \kappa_{icai}^{*}}{\left(\epsilon_{b} + \epsilon_{c} - 2\epsilon_{i}\right) \left(\epsilon_{a} + \epsilon_{c} - 2\epsilon_{i}\right)},
\end{equation}
where $N_{\rm occ}$ is the number of occupied canonical orbitals, $N_{\rm virt}$ is the number of virtual canonical orbitals, and $\epsilon_{i/a/b/c}$ are eigenvalues of HF states.  
This matrix can be calculated using $O(N_{\rm pw}N_{\rm occ}N_{\rm virt}^2)$ operations and $O(N_{\rm pw}N_{\rm occ}N_{\rm virt})$ memory in the plane-wave basis set. In the first step, $N_{\rm virt}$ is chosen as large as possible ($N_{\rm occ} + N_{\rm virt} = N_{\rm pw}$). Assuming natural orbitals are ordered according to eigenvalues of the approximate MP2 density matrix, we then choose first several natural orbitals (the exact number for each specific case described in the text) as virtual orbitals for subsequent calculations.

\section{Quantum Error Correction}\label{app:error correction}
The algorithms presented here require coherence times and noise levels that are beyond the reach of any qubit technology, and require fully fault-tolerant quantum computation.
To achieve this we use Quantum Error Correction (QEC), where many physical qubits are used to encode a single logical qubit.
In~\cite[Section 4]{Blunt2022a} we detail the error correction scheme we use, which is based on the surface code on a 2D grid, following the scheme given in ``A Game of Surface Codes"~\cite{Litinski2019}.

We choose a target failure probability of $p_\text{fail}=1\%$ for a full execution of the quantum algorithm, which we divide into error budgets $p_\text{fail}^\text{log}=0.9\%$ for logical errors, and $p_\text{fail}^\text{MSD}=0.1\%$ for undetected errors in magic state distillation.

Due to the large number of Toffoli gates required for this algorithm, we utilise gate synthillation in magic state factories that produce CCZ states \cite{Jones2013a, Eastin2013, Campbell2017a, Gidney2019a,Litinski2019d} instead of $T$ factories.
Gate synthillation allows us to produce $\ket{\text{CCZ}}$ states with a lower overhead than producing four $T$ states to implement a Toffoli gate.
For the largest systems considered here, we use the (15-to-1)${}^4_{9,3,3}$ $\times$ (8-to-CZZ)${}_{25,9,9}$ factory from \cite{Litinski2019d}. 
It has a sufficiently low failure probability, below the target error $p_\text{fail}^\text{MSD}/N_\text{Toff}$ for the largest system; we can use smaller factories for smaller instances.
A smaller code distance than $d$ in the logical computation can be used for the magic state factory to reduce its footprint and runtime. In fact, rectangular code patches with distinct distances for X, Z, and time (as indicated by the subscripts) can be used as the factory is more prone to some types of error than others.

Like in~\cite{Blunt2022a} we assume the computation proceeds as fast as consuming one magic state qubit per logical clock cycle, where a logical clock cycle is equivalent to $d$ code cycles, and $d$ is the code distance. Consequently, we ensure that the number of magic state factories available is high enough that a single magic state is available every three logical cycles, which typically requires multiple magic state factories.
The length of the computation is $3N_\text{Toff}$ logical cycles.

The logical error budget bounds the allowed logical failure probability per logical cycle, which is given by the Fowler-Devitt-Jones formula~\cite{Fowler2012}. Hence the computational code distance $d$ must be chosen such that
\begin{equation}
A \left (\frac{p}{p_{\textrm{thr}}} \right )^{\frac{d + 1}{2}} \le \frac{p_\text{fail}^\text{log}}{3 N_\text{Toff} N_{\textrm{qubits}}} ,
  \label{eq:fowler_formula}
\end{equation}
where $p$ is the probability of a physical error, $p_\text{thr}\approx0.01$ is the threshold of the surface code, and $A=0.1$ is a numerically determined constant.

In order to estimate the physical resources, we model a 2D superconducting device as in \cite{Blunt2022a}, with an error rate one order of magnitude better than current superconducting devices~\cite{google_quantum_ai_suppressing_2023, Krinner2022b}, i.e. $p  = 0.01 \%$.
This allows us to solve Eq.~\ref{eq:fowler_formula} for $d$.
The total number of logical qubits $n_{L}$ is given by the number of qubits required by the algorithm and the routing required by the \textit{fast-block} layout~\cite[Figure 13]{Litinski2019}.
The total number of physical qubits is then $(2d^2 -1)n_L$ for algorithm and routing in the rotated surface code, together with those required by the magic state factories.

\section{Block encoding of a squared matrix}
\label{app: square}
A main feature of the double factorised Hamiltonian is that it contains terms that are a square $A^2$ of another expression, see Eq.~\ref{eq:full_hamiltonian}. Starting from a block encoding of $A/\alpha$ (with subnormalisation $\alpha$), in principle the square can be implemented by multiplying block-encoded matrices \cite{gilyen2019quantum,von_burg_quantum_2021,sunderhauf2023generalized}, yielding a block encoding $\frac{1}{\alpha^2} A^2$.
Instead, when the double factorisation algorithm was conceived in \cite{von_burg_quantum_2021}, the square was implemented by applying the second Chebyshev polynomial
\begin{equation}
    T_2(x) = 2x^2-1
\end{equation}
to $A/\alpha$. This results in a block encoding of $\frac{2}{\alpha^2}A^2-1$, whose constant shift can be computed classically. Compared to multiplication, the subnormalisation of the Chebyshev polynomial is better by a factor of 2, at the same query complexity. This factor of two has been taken into account when computing the subnormalisation Eq.~\ref{eq:subnormalisation}.

Chebyshev polynomials can be implemented via qubitisation \cite{low_hamiltonian_2019}. Here we explicitly demonstrate $T_2(x)$.
Let
\begin{equation}
    U = \begin{pmatrix}A/\alpha & B \\ B^\dag & C\end{pmatrix},\ \mathcal R = \begin{pmatrix}\mathbb{1} & \\ & -\mathbb{1}\end{pmatrix}
\end{equation}
be a Hermitian block encoding of the Hermitian $A/\alpha$, and the reflection around the coding subspace. Due to unitarity we have $(A/\alpha)^2 + BB^\dag = \mathbb{1}$, such that
\begin{multline}    
    U \mathcal R U = \begin{pmatrix}A/\alpha & B \\ B^\dag & C\end{pmatrix}\begin{pmatrix}A/\alpha & B \\ -B^\dag & -C\end{pmatrix}
    = \begin{pmatrix}
        (A/\alpha)^2 - BB^\dag & * \\
        * & *
    \end{pmatrix}
    = \\
    \begin{pmatrix}
        (A/\alpha)^2 + (A/\alpha)^2 - \mathbb{1} & * \\
        * & *
    \end{pmatrix}
    = \begin{pmatrix}
    2(A/\alpha)^2 -\mathbb{1} & * \\ * & *\end{pmatrix},
\end{multline}
where we have omitted calculation of the junk blocks of the final block encoding.
The reflection $\mathcal{R}$ must implement a $-1$ phase outside of the coding subspace, otherwise $-T_2(x)$ is implemented. To this end, we have added a CZ to the implementation of the reflection in Fig.~\ref{fig:block_encoding}.

\bibliography{main.bib}
\end{document}